%% file: main.tex
\documentclass[conference]{IEEEtran}
\IEEEoverridecommandlockouts
\usepackage{savesym}
\savesymbol{theoremstyle}
\savesymbol{algorithm}

\usepackage{graphicx} 
\usepackage{listings}
\usepackage[algo2e,linesnumbered, ruled,vlined]{algorithm2e}
\usepackage{amssymb,amsfonts}%
\usepackage{amsthm}
\usepackage{mathtools}

\usepackage[font=small]{caption}
\usepackage{verbatim}
\usepackage{epsfig}
\usepackage{float}
\usepackage{forest}
\usepackage{subfig}
\usepackage{setspace}
\usepackage{xspace}
\usepackage{macros}

\newcommand*{\Scale}[2][4]{\scalebox{#1}{$#2$}}%
\begin{document}

\title{Fast Robust Monitoring for Signal Temporal Logic with Value Freezing Operators (\STLstar)} 
\author{\IEEEauthorblockN{Bassem Ghorbel}
\IEEEauthorblockA{Colorado State University \\
USA\\
bassem@colostate.edu}
\and
\IEEEauthorblockN{Vinayak S. Prabhu}
\IEEEauthorblockA{Colorado State University \\
USA\\
vinayak.prabhu@colostate.edu}
}
\maketitle

\setlength{\abovedisplayskip}{2pt}
\setlength{\belowdisplayskip}{2pt}

\begin{abstract}    
\input{abstract}
\end{abstract}

\input{intro}

\input{stlstar}

\input{expressiveness}

\input{ast}

\input{algo}

\input{example}

\input{robust}

\input{complexity}

\input{exp}

\input{conclusion}

\section*{Acknowledgment}
  This work was supported in part by the National Science Foundation by a CAREER award
  (grant number 2240126).

\bibliographystyle{ieeetr} 
\bibliography{stlstar}

\clearpage
\input{appendix}

\end{document}

%% file: abstract.tex
Researchers have previously proposed augmenting Signal Temporal Logic (STL) with the value freezing operator in order to express engineering properties that cannot be expressed in STL. This augmented logic is known as \STLstar.
The previous algorithms for \STLstar monitoring were intractable, and did not scale
formulae with nested freeze variables.
We present offline discrete-time monitoring algorithms with an acceleration heuristic, both for
Boolean monitoring as well as for quantitative robustness monitoring.
The acceleration heuristic operates over time
intervals where subformulae hold true, rather than over the original  trace sample-points. We present experimental validation of our algorithms, 
the results show that our algorithms can monitor over long traces for formulae with two or three nested freeze variables.
Our work is the first work with monitoring algorithm implementations for \STLstar formulae with nested freeze variables.

%% file: intro.tex
\section{Introduction}

In the context of Cyber-Physical Systems (CPS) and Control, \emph{Signal Temporal Logic} (STL) has found wide adoption as a trace property specification formalism~\cite{FainekosSUY12,RamanDSMS15,RamanDMMSS17,DeshmukhHJMP17,Sankaranarayanan17,BartocciDDFMNS18,KongJB17,ErnstSZH21,LiuMB22}.
\STL, which can be seen as a flavor of Metric Temporal Logic (MTL)~\cite{Koymans}, allows specification of properties such as ``if the temperature rises above 100 degree Celsius at any point in time, then it will within 5 time units, fall below 50 degree Celsius, and stay below 50 degree Celsius for at least 2 time units''.
Such specifications incorporate both a temporal aspect (e.g., ``within 5 time units'') as well as a signal constraints aspect (e.g., ``temperature $>$ 100'').
Two key problems over temporal logics for CPS are
(1)~\emph{Boolean monitoring}: checking whether a given trace satisfies a temporal logic specification; and
(2)~\emph{Robustness monitoring}: defining a quantitative measure of how well a given trace satisfies a temporal logic specification, and computing this numerical value.
Tools such as 
STaLiRo, Breach,  FALSTAR, and FalCAuN use such robustness monitoring procedures for \STL
for test generation in order to falsify \STL
specifications~\cite{FainekosSUY12,DeshmukhDGJJS17,ErnstSZH21,Waga20}.

While  \STL demonstrates the utility of  temporal logics for verification, control, and testing of
CPS, it is unable to express commonly occurring properties in biological and engineering systems, such as oscillatory properties, as has been noted by researchers~\cite{Brim14,Bakhir19,SilvettiNBB18}.
A natural manner to increase the expressive power of \STL is to add \emph{freeze quantification}, which allows the capture of a signal value into a freeze variable, to be used later in the trace for comparison~\cite{Brim14}.
Freeze quantification was first introduced in the context of temporal operators as
\emph{time} freeze quantification in~\cite{AlurH94}, and  the resulting  increase in expressive power was proved in~\cite{BouyerCM10}.
The work
\cite{Brim14} introduced the logic \STLstar and showed how \emph{value} freeze quantification in
the \emph{signal domain} enabled specification of properties which are believed to be outside of the scope of STL.
We illustrate the value freeze operator. Consider the requirement:
\emph{``At some future time (during interval $I_1$), there is a local maximum over 5s, then at another future time (during interval $I_2$), there is a local minimum  over 5s''}.
This requirement can be written in \STLstar as
$\varphi_0=\Diamond_{I_1} s_{*1}. \big(\left(\Box_{[0,5]} s \le s_{*1}\right)\, \wedge\, \Diamond_{I_2} s_{*2}. \left(\Box_{[0,5]} s \ge s_{*2}\right) \big)$.
The freeze variable $s_{*1}$ freezes the signal value at some point in interval $I_1$, and this frozen value is accessed as $ s_{*1} $ for the local maxima check in $\Box_{[0,5]} s \le s_{*1} $, and similarly for freeze variable $s_{*2}$.

However, the increased expressivity of freeze quantifiers incurs a price on
monitoring algorithms --- which is to be expected since the monitoring problem for
 temporal freeze quantifier augmented MTL is PSPACE  hard~\cite{AlurH94,BouyerCM10}.
An alternative mechanism to increase temporal logic expressivity is by adding  \emph{first order} quantification~\cite{Basin15}; however first order quantification similarly makes monitoring intractable in the general case~\cite{BakhirkinFHN18}.
Orthogonal to freeze quantification, monitoring algorithms also depend on whether
pointwise semantics of traces (a trace being a sequence of signal values) or whether continuous time semantics (with traces being completed using linear or piecewise constant interpolation from sampled values) are used; the resulting impact on
algorithm intricacy  can be seen even in \STL~\cite{DonzeFM13,DeshmukhDGJJS17}.
The Boolean monitoring algorithm in~\cite{Brim14} for \STLstar  uses continuous time semantics with linear interpolation; it is complex, and involves manipulation of polygons. Even in the case of a single value
freeze operator, and even for \emph{approximate} monitoring in an attempt to make the problem tractable, the algorithm  in~\cite{Brim14} remains complex.
Due to the complicated nature of the algorithm which involves manipulating polygons,  \cite{Brim14} did not obtain a precise complexity bound: it was only shown that ``steps of the algorithm has at most polynomial complexity to the number of polygons and the number of
polygons grows at most polynomially in each of the steps''. Their monitoring
experiments showed scalability limitations --
over an hour of running time for signals containing 100 time-points, and even restricted to  \STLstar formulae containing only one active freeze variable.

The work of~\cite{Brim13} proposes a \STLstar robustness monitoring algorithm for pointwise semantics, but the algorithm calls -- for every possible binding of a freeze operator to a frozen
signal value --   a subroutine which has
 a $O(|\pi|^{2})$ dependence where
 $|\pi|$ is the trace length. If $|V|$ is the number of freeze quantifiers used, there are
 $|\pi|^{|V|}$ freeze operator bindings,  resulting in an overall algorithm time complexity of
 $O(|\pi|^{|V|+2})$,
 and hence the procedure is  not tractable even for one freeze variable.
The experiments in~\cite{Brim13} are over specialized formulae: only one freeze variable, no until operator, and no nested temporal operators within the scope of the freeze variable; this allows the use of a special algorithm, which is not described.
The recent work of~\cite{Ghorbel24}  examined the Boolean monitoring problem for the
fragment of \STLstar in which any subformula could contain at most one ``active'' freeze variable (i.e., no nested freeze variables). For this one variable fragment of \STLstar,  they presented an efficient Boolean monitoring algorithm in the pointwise semantics scaling to trace lengths of 100k. That work did not address the robustness monitoring problem.
In the present work, we build upon~\cite{Ghorbel24}, lift the one variable limitation, and examine the Boolean as well as the robustness  monitoring problem for the
full logic of \STLstar with nested value freeze variables.

\noindent\textbf{Our Contributions.}
Our main contributions are:\\
(\emph{I}) We investigate which engineering properties require more than one active value freeze variable, and we present natural requirements using two and three freeze variables which we conjecture cannot be  specified in one variable \STLstar.\\
(\emph{II}) We present offline  Boolean monitoring algorithms for full \STLstar in the pointwise semantics, for both uniformly, and non-uniformly time-sampled traces.
Our algorithms use an \emph{acceleration} technique based on the work from~\cite{Ghorbel24}. We show that the acceleration technique can be used for Boolean monitoring for full
\STLstar.
Suppose we have $|V|$ freeze variables in a formula $\varphi$. In order to check whether a trace $\pi$
satisfies the formula, we need to iterate over the trace, and at each trace point, there are $|\pi|^{|V|}$ possibilities for the bindings of the $|V|$ freeze variables as each freeze variable can be bound to the signal value at any sample point. This gives us a search space of approximately
$|\pi|^{|V|+1}$ for monitoring.
We show that we need \emph{not} iterate over all  timepoints in the trace for each subformula and for every freeze variable binding: it turns out that in most cases we can iterate over \emph{intervals} rather than individual timepoints -- these intervals are the partitions of the time-stamps where the relevant subformulae have a true value throughout.
This results in a monitoring complexity of $O\bigl(|\pi|^{|V|} \cdot \max\bigr(\log(|\pi|), \, |\varphi| \!\cdot\! |\intvl(\varphi)|\bigr)\bigr)$
for a uniformly sampled trace in practice, where $|\intvl(\varphi)|)$ is the resulting
number of such  intervals for  any sub-formula of $\varphi$ (and similarly for  non-uniformly sampled traces).
The acceleration technique leverages the idea that these intervals do not change much from one freeze binding to the next for realistic traces where signal values do not vary wildly from one timepoint to the next.
In practice, $|\intvl(\varphi)| << |\pi|$. Thus, our acceleration heuristic reduces the exponent in $|\pi|^{|V|+1}$ by one (at the expense of $\log(|\pi|)$).\\
(\emph{III}) We present offline robustness monitoring algorithms for full \STLstar.
Unfortunately, the intervals idea of Boolean monitoring cannot directly be used for value computation, as from  freeze binding to one sampled signal value to the next, the quantitative robustness \emph{value} would indeed change. However, we show that the acceleration heuristic can be used for the robustness \emph{decision problem} which asks whether the robustness value is less than or equal to some given threshold. 
We achieve the  robustness value  computation   by binary search over a conservative robustness range.
We show how to compute this conservative range given a formula and a trace.
We also give a non-interval monitoring algorithm, improving upon the algorithm
from~\cite{Brim13} by a careful handing of the until operator; this improves
the $|\pi|^{|V|+2}$ time complexity factor from~\cite{Brim13} to $|\pi|^{|V|+1}$.\\
(\emph{IV}) We obtain time complexity bounds for our algorithms.\\
(\emph{V}) We implement our algorithms and present experimental results. We show that with the accelerated algorithms, monitoring for two nested freeze variables remains tractable -- 
Boolean monitoring for traces of size 10k takes about 3 minutes, and  about 17 minutes for robustness value computation (with the final robustness value estimate being within 2\% of the actual value). We believe two nested freeze variables suffice to capture most engineering properties of interest in \STLstar.
We are also able to monitor for three nested freeze variables over traces of size 500 in about 2.5 minutes, and for traces of size 1k in about 22 minutes for Boolean monitoring.
We note that~\cite{Brim14,Brim13} did not implement  monitoring for two freeze variables in their experiments  as their algorithms do not scale to two freeze variables.
Thus, ours is  the \emph{first} work which presents implemented monitoring algorithms for \STLstar formulae with nested freeze variables, both for Boolean monitoring as well as for robustness value computation.


\smallskip\noindent\textbf{Related Work.}
Logics augmented with frequency constructs have recently  been proposed  for property specification in the frequency domain~\cite{DonzeMBNGS12,Johnson17}.
Freeze quantification enables expanded expression of properties in the time domain~\cite{Parache21}.
In general, it is well known that both time and frequency domain analyses provide useful information about signals.
Efficient algorithms for \emph{time} freeze quantification  are presented in~\cite{DokhanchiHTF16,GhorbelP22,Ghorbel23}.

%% file: stlstar.tex
\section{Value-freezing Signal Temporal Logic}

\noindent\textbf{Signals/Traces.}
Let $\dm \in \nat_{>0}$, a $\reals^{\dm}$ valued \emph{signal} or a \emph{trace} is a
pair $(\sigma, \tau)$, where $\sigma = \sigma_0, \sigma_1, \ldots,  \sigma_{|\pi| -1}$ is a finite sequence of elements from $\reals^{\dm}$, and $\tau = \tau_0, \tau_1, \ldots,  \tau_{|\pi| -1}$ are the corresponding timestamps from $\reals_+$. The signal value at timestamp $\tau_i$ is $\sigma_i\in \reals^{\dm}$ and $i$ is a position index. The $k$-th signal dimension of $\sigma =\tuple{a^1, \ldots, a^{\dm}}$, namely $a^k$, is denoted $\sigma^k=\sigma_0^k,\sigma_1^k, \dots, \sigma_{|\pi|-1}^k$. 
We require the times to be monotonically increasing, that is $\tau_i < \tau_{i+1}$ for all $i$.
If $\tau_i = i\cdot\Delta$ for some $\Delta>0$, the traces are said to be \emph{uniformly sampled}; \emph{non-uniformly sampled} otherwise.

To reduce notation clutter, we use $\overline{x}=x^1,x^2,\dots,x^{\dm}$ for any variable $x$.

%

\begin{definition}[\STLstar Syntax]
  \label{def:Syntax}
  Given a signal arity $\dm$,  and a finite set of \emph{freeze variables} $\set{s^1_*, \ldots, s^{\dm}_*}$
  for each signal dimension $1\leq k\leq \dm$,
  the syntax of value-freezing signal temporal logic (\STLstar) is defined as follows:
  \begin{itemize}
        {\setstretch{1.2}
        \item  $f(\overline{s}) \sim 0$;\ and \quad
          $f_1(\overline{s}) \sim f_2(\overline{s_*}) $.

  \item  $ \neg \varphi $; and $   \varphi_1  \wedge     \varphi_2 $; and  $   \varphi_1  \vee    \varphi_2 $.
    
  \item $\Box_I \varphi$; and $ \Diamond_I \varphi$; and $ \varphi_1 \until_I \varphi_2$.
  \item $s^k_*. \varphi$.
    }
  \end{itemize}    
  where $s^k \in \set{s^1, \ldots, s^{\dm}}$  is a \emph{signal  variable} ($s^k$ refers to the $k$-th signal dimension), $f(\overline{s}) \sim 0$ are signal predicates, $f_1(\overline{s}) \sim f_2(\overline{s_*})$ are signal constraints, $f$, $f_1$ and $f_2$ are arbitrary functions, and $I=[a,b]$ is an interval where $a$ and $b$ are positive reals, 
  $s^k_{*}$ is a \emph{freeze variable} corresponding to the signal-value freeze operator  $``s^k_*."$, and $\sim \in \set{ <, >, \leq, \geq} $ is the standard comparison operator. 

 The freeze operator ``$s^k_*.$'' binds the current $k$-th signal dimension value to the frozen value $s^k_{*}$ in the signal constraints.\qed
  \end{definition}

  The original work in~\cite{Brim14} allowed only affine functions for $f,f_1$ and $ f_2$; but as our treatment is in discrete-time, we can handle arbitrary computable functions.

  \textbf{Note:} We can freeze the same signal dimension multiple times in an \STLstar formula. In that case, we use a subscript $h \in \mathbb N_{> 0}$ to indicate which frozen value refers to which signal-value freeze operator.
Thus, $s_{*h_1}^k$ and $s_{*h_2}^k$ for $h_1 \neq h_2$ are considered different freeze variables.
When we freeze a signal dimension only once in a \STLstar formula, we omit the superscript $h=1$ (check example \ref{runningex}).



\begin{definition}[Semantics]
  \label{def:Semantics}
  Let $\pi = (\sigma_0, \tau_0),  $ $ (\sigma_1, \tau_1), \ldots, $ $(\sigma_{|\pi|-1}, \tau_{|\pi|-1})$ be a finite
  timed signal of arity $\dm$.
  For a given environment $\env: V \rightarrow \reals$ binding freeze variables to 
  signal values, and position index $0\leq i \leq  |\pi|-1$, the satisfaction
  relation $(\pi, i, \env) \models \varphi$ for an \STLstar formula $\varphi$ of arity $\dm$
  (with freeze variables in $V$) is defined as
  follows.
  \begin{itemize}
    {
     \setstretch{1.2}
  \item $  (\pi,i ,\env)  \models f(\overline{s}) \sim 0 $ iff $f(\overline{\sigma_i}) \sim 0$.

    \item $  (\pi,i ,\env) \models \neg \varphi $ iff $(\pi,i ,\env) \not \models \varphi$.

    \item $  (\pi,i ,\env) \models \varphi_1
    \begin{smallmatrix}\vee\\ \wedge
    \end{smallmatrix} \varphi_2 $ iff $(\pi,i ,\env) \models \varphi_1 \begin{smallmatrix}\text{or}\\ \text{and}
    \end{smallmatrix} (\pi,i ,\env) \models \varphi_2$.

  \item $\Scale[0.97]{  (\pi,i ,\env) \models \Diamond_{[a,b]} \varphi$ iff $\exists j$, $\tau_j \in \tau_i  +  [a,b]$, s.t. $(\pi,j ,\env)  \models \varphi}$.
    \item $  (\pi,i ,\env)   \models    \Box_{[a,b]} \varphi$ iff $\forall  j$, $\tau_j \in \tau_i + [a,b]$, we have  $(\pi,j ,\env)  \models \varphi$.

        \item $  (\pi,i ,\env) \models \varphi_1 \until_{[a,b]} \varphi_2 $ iff $(\pi,j ,\env)$ $ \models\varphi_2$ for some $j \ge i$ with $\tau_j \in \tau_i + [a,b]$  and $ (\pi,k ,\env) \models \varphi_1$ $\forall i \le k<j$.

    \item $  (\pi,i ,\env) \models s^k_*. \varphi$ iff $(\pi,i ,\env[s^k_*:= \sigma_i^k])  \models \varphi$;
    where $\env[s^k_*:= \sigma_i^k]$ denotes the environment $\env'$ defined as $\env'(x) = \env(x) $
    for $x \neq s^k_*$, and $\env'(s^k_*) = \sigma_i^k$.
    \vspace*{2mm}
    \item $  (\pi,i ,\env) \models \Scale[0.93]{f_1(\overline{s}) \sim f_2(\overline{s_*})$ iff $f_1(\overline{\sigma_i}) \sim f_2(\env(s^{1}_*),..,\env(s^{\dm}_*))}$.\hspace*{-10mm}
  }  
  \end{itemize}
We say trace $\pi$ satisfies \STLstar formula $\varphi$ if
  $(\pi, 0, \zeroenv) \models \varphi$ where $\zeroenv$ denotes the freeze environment where  all freeze variables $s^k_*$ are mapped to their  corresponding $\sigma_0^k$ values.
  \qed
  \end{definition}

\begin{example}
We consider a single dimension signal $s$ and we freeze it twice ($h=1$ and $h=2$) in the following formula:
$\varphi_0=\Diamond_{I_1} s_{*1}. \bigl(\Box_{[0,5]} s \le s_{*1} \wedge \Diamond_{I_2} s_{*2}. (\Box_{[0,5]} s \ge s_{*2})\bigr)$.
The requirement of $\varphi_0$ is: \emph{``at some time in the future (during $I_1$), there is a local
  maximum over 5s, then at another time in the future (during $I_2$), there is a local minimum over 5s ''}.
\qed
\end{example}

To reduce notation clutter, we will simply write $(i ,\env)$ instead of $(\pi,i ,\env)$ in the remainder of this paper since for any given \STLstar formula, we will be using the same trace $\pi$.
We use the phrase \emph{$i^{th}$ instantiation} of a freeze variable $s^k_*$ to mean the environment $\env$ where the freeze variable $s^k_*$ is assigned the value $\sigma_i^k$.

%% file: expressiveness.tex
\section{Expressiveness of \STLstar} \label{expressiveness}
Authors in \cite{Brim14} introduced \STLstar with multiple freeze variables but did not give any requirement with more than one freeze variable.
In this section, we present some interesting specifications that require more than just a single freeze variable. We provide 
four engineering properties that require \STLstar expressiveness with nested freeze variables. 
\begin{example} [Running Example]
\label{runningex}
Eventually $e_1$ happens and after that, eventually, $e_2$ happens and 2-time units after that, the values of $s$ are always within 20 $\%$ of the average of the value of $s$ when $e_1$ happened and the value of $s$ when $e_2$ happened (Stabilization of $s$ around not known multiple values in advance). $e_1$ and $e_2$ can be any signal predicates: $\varphi_1 =$ 
    \[\Scale[0.85]{\Diamond \biggl(e_1 \wedge s_{*1}.\Bigl(\Diamond \bigl(e_2 \wedge s_{*2}. \Box_{[2,T]} s \in [0.8\dfrac {s_{*1} + s_{*2}}{2},
        1.2\dfrac {s_{*1} + s_{*2}}{2}  ]\bigr)\! \Bigr)\! \biggr). \qed}\]
    
\end{example}
\begin{example}
Eventually the value of $s^1$ is greater than 5 (at an unknown moment $t_1$) and after that, eventually the value of $s^1$ is greater than 10 (at an unknown moment $t_2$), and after that, eventually the value of $s^2$ is greater than the value of $s^2$ at $t_1$ plus the value of $s^2$ at $t_2$ until $s^1$ is less than 5:
    $\varphi_2 =$ 
    \[\Scale[0.85]{\Diamond\biggr( s^1 > 5 \wedge s^2_{*1}.\Diamond \Bigl(s^1 > 10 \wedge s^2_{*2}.\Diamond\bigl( (s^2 > s^2_{*1} + s^2_{*2}) \until s^1 < 5\bigr) \Bigr)\biggr).\qed}\]
\end{example}

    Previous \STLstar work used only linear constraints (and predicates) of the form $\sum a_i s^i + \sum b_i s^i_*  \sim r$ where $a_i$, $b_i$ and $r$ are constants, while we do not impose such restrictions and we are free to use any type of functions.

\begin{figure}[h]%
    \centering
    \subfloat[Signal that satisfies $\varphi_3$ and $\psi$]{{\hspace*{-5mm}\includegraphics[scale=0.25]{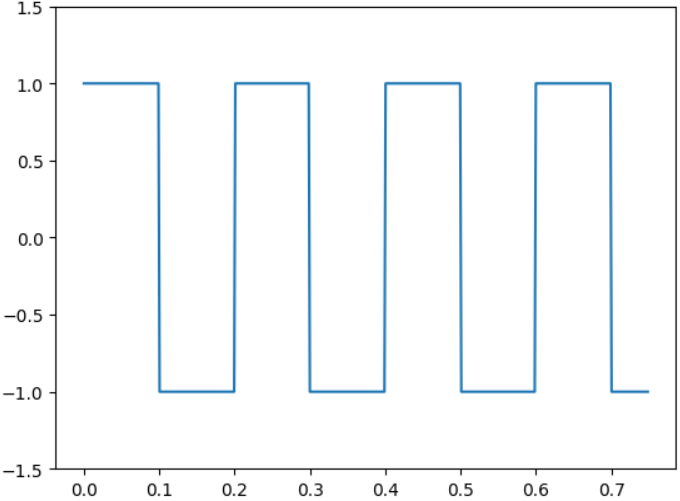} }}%
    \
    \subfloat[Signal that satisfies $\varphi_3$, not $\psi$]{{\includegraphics[scale=0.235]{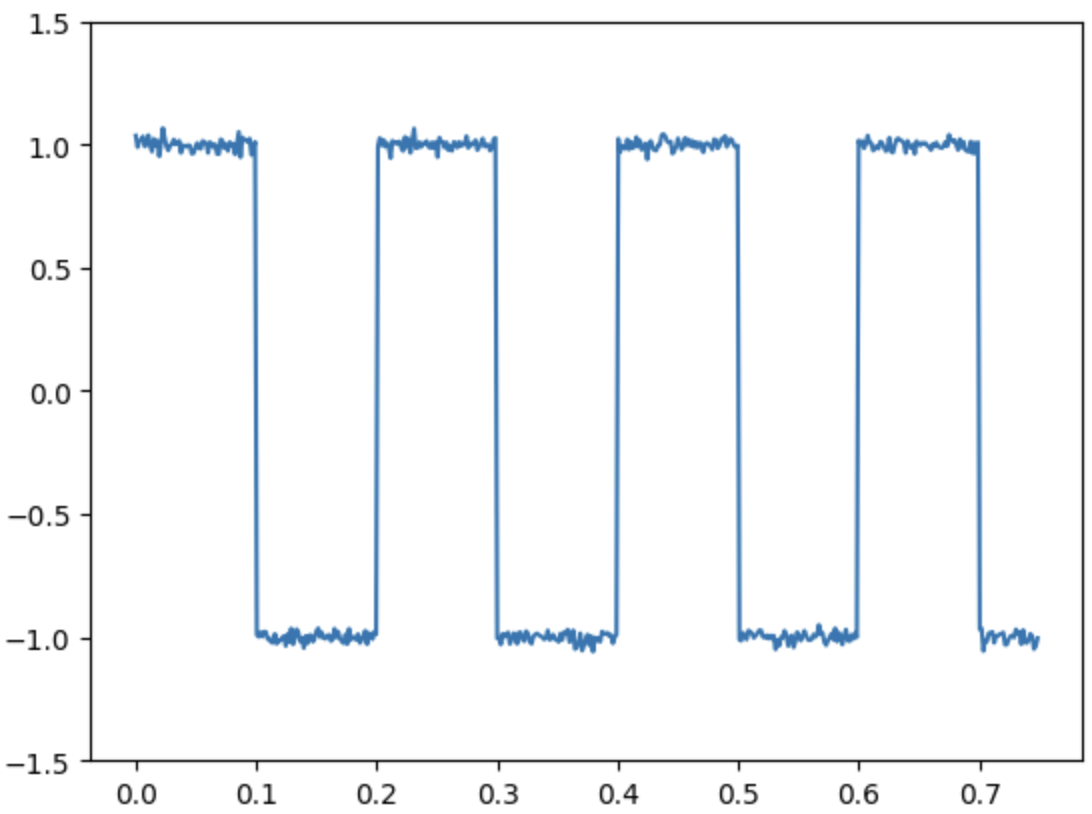} }\hspace*{-5mm} }
    \caption{Signals that satisfy the rectangular pulse requirement}
    \label{fig:signals}%
\end{figure}

\begin{figure}[h]
    \centering
    \includegraphics[scale=0.23]{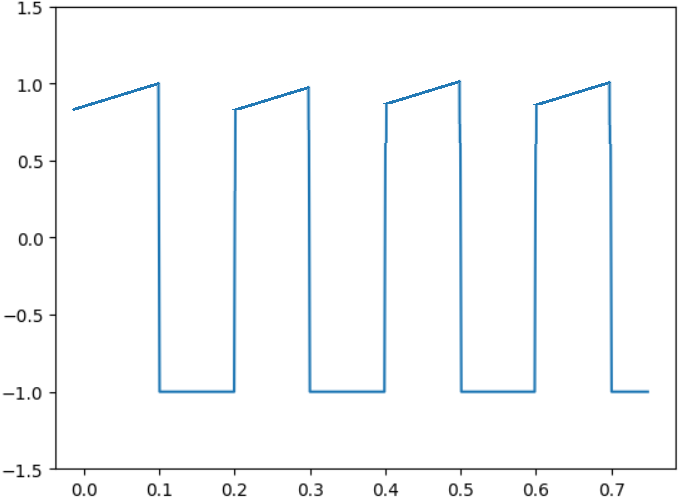}
    \caption{Signal that can be accepted by $\psi$ with large $\epsilon$}
    \label{fig:problem}
\end{figure}

\begin{example}
    Check if $s$ is a rectangular pulse signal with unknown pulse value: $\varphi_3=$ 
    \[\Scale[0.83]{\Box s_{*1}.\Bigl(|s_{*1}-s| \le \epsilon \text{ } \until \bigl(|s_{*1}-s| \ge \Delta \wedge s_{*2}.(|s_{*2}-s| \le \epsilon \text{ } \until |s_{*1}-s| \le \epsilon )\bigr)\Bigr)}\] 
    where $\epsilon$ is an error threshold and $\Delta$ is the minimum pulse amplitude. We note that the $\epsilon$ value can be in function of the frozen values, for example, $\epsilon=0.1 \times s_{*1}$.

To  understand the logic behind the above formula $\varphi_3$, we  look at Figure \ref{fig:signals} (a): At any given time point (which is represented by the $\Box$ at the beginning), the first freeze variable $s_{*1}$, freezes the value 1 (or -1) and the future values of $s$ must remain within that frozen value $s_{*1}$ with an $\epsilon$ error \emph{until} a sudden increase or decrease of the amplitude of $s$ by at least $\Delta$. When that happens, the second freeze variable $s_{2*}$ freezes the current $s$ value which is -1 (or 1) and the future values of $s$ must remain within that frozen value $s_{*2}$ with an $\epsilon$ error \emph{until} $s$ is equal to the first frozen value $s_{*1}$ with an $\epsilon$ error.

Some would suggest that this requirement can be expressed using just $\STLstar_1$ as follows
\[\Scale[0.9]{\psi= \Box s.\Bigl(|s_{*}-s| \le \epsilon \text{ } \until \bigl(|s_{*}-s| \ge \Delta \wedge( |s'| \le \epsilon \text{ } \until \text{ } |s_{*}-s| \le \epsilon )\bigr)\Bigr)}\]
where $s'$ is the derivative of $s$.
However, this formula would have problems for signals with noise. Using the derivative makes our formula look at the signal locally, and will not give an idea on how the signal behaves globally (Figure \ref{fig:signals}). 

In addition, for a larger$\epsilon$ threshold, $\psi$ can end up accepting signals that should not be accepted, 
see Figure \ref{fig:problem}.\qed
\end{example}
\begin{example}
    Check if a signal $s$ follows a repeating two stairs signal with unknown pulse values: $\varphi_4=$\\
    $\Scale[0.84]{\Box s_{*1}.\Biggl(|s_{*1}-s| \le \epsilon \text{ } \until \biggl(|s_{*1}-s| \ge \Delta \wedge s_{*2}.\Bigl(|s_{*2}-s| \le \epsilon \text{ } \until \bigl(|s_{*2}-s| \ge \Delta}$ \\[-2mm]
    \hspace*{20mm}$\Scale[0.84]{ \wedge \text{ } s_{*3}.(|s_{*3}-s| \le \epsilon \text{ } \until \text{ } |s_{*1}-s| \le \epsilon)\bigr)\Bigr)\biggr)\Biggr)}$.\qed 
\end{example}


%% file: ast.tex
\section{\STLstar Syntax Trees} \label{sec-syntaxtree}
Each \STLstar formula has a corresponding syntax tree that depicts the hierarchical syntactic structure of the formula. Our monitoring procedure will depend on this syntax tree.

The basic structure over which our algorithm will operate will be subtrees corresponding to various freeze variables. The following example explains subtrees, parents and roots.

\begin{figure}[h]
\centering
\includegraphics[trim=0  0 0 20, scale=0.7]{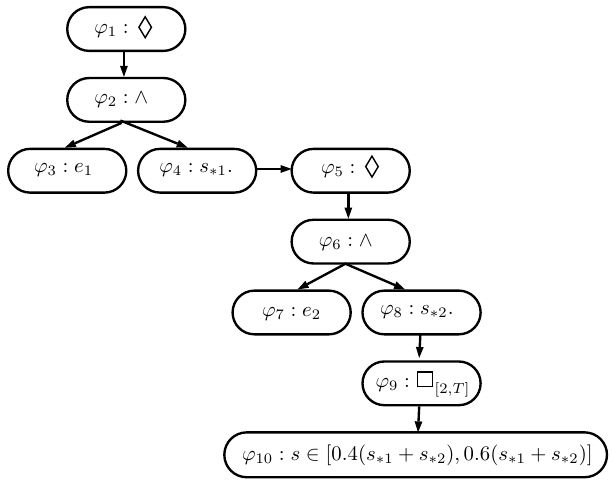} 
\caption{Syntax Tree for Example \ref{runningex}}
\label{fig:Tree}
\end{figure}

\begin{example} 
  \label{example:SubTrees}
  Consider formula $\varphi_1$  from Example \ref{runningex}. Figure \ref{fig:Tree} has
  its associated syntax tree.
  The formula subscripts correspond to a reverse topological sort of the syntax tree. The freeze variable ordering given by a reverse topological sort is
  $s_{*2} <_{\revtop} s_{*1}$. The nodes in $\subtree_{\varphi_0}(s_{*2})$ are $\set{\varphi_9,\varphi_{10}}$. The nodes in $\subtree_{\varphi_0}(s_{*1})$ are $\set{\varphi_{5},\varphi_{6},   \varphi_{7},\varphi_8}  $; and in $\tsubtree(\varphi)$ are $\set{\varphi_1,\varphi_2,\varphi_3,\varphi_4}  $. $\subtree_{\varphi_0}(s_{*2}).\roottree= \varphi_9$, $\subtree_{\varphi_0}(s_{*2}).\parent= \varphi_8$, $\subtree_{\varphi_0}(s_{*1}).\roottree = \varphi_5$, $\subtree_{\varphi_0}(s_{*1}).\parent = \varphi_4$.
  \qed
\end{example}

%% file: algo.tex
\section{\STLstar Boolean Monitoring Algorithm} \label{monitor}
The monitoring problem we consider in the paper is as follows:
Given a trace $\pi$, and an \STLstar formula $\varphi$,
do we have  $(\pi, 0, \zeroenv) \models \varphi$?
Before presenting our algorithm, we need first some notations.\\
\textbf{Notation:} For any \STLstar subformula $\psi$, we note:
\begin{itemize}
    \item $\point(\psi)$: a list of length $|\pi|$ of $\true$ and $\false$ values where for each $0\le i \le |\pi|-1$, each value represents whether $(i ,\env)\! \models \psi$ or not for a given environment $\env$.
    \item $\intvl(\psi)$: a list of intervals $[\tau_{a_1},\tau_{a_2}], [\tau_{b_1},\tau_{b_2}],$ $ \dots, [\tau_{z_1},\tau_{z_2}]$, where $\tau_{a_1}, \tau_{a_2} \dots \tau_{z_2}$ are timestamps and $a_1 \le a_2 < b_1 \le b_2 \dots <z_1 \le z_2$, each interval represents a sequence of $\true$'s appearing in $\point(\psi)$.
\end{itemize}
\begin{example}
Suppose we have the following trace $(\sigma_i,\tau_i), i \in[0,10]$: (5,0), (3,1), (7,2), (-2,3), (-5,4), (3,5), (-1,6), (3,7), (4,8), (5,9), (6,10) and the subformula $\psi=s \ge 0$, then:
\begin{itemize}
    \item $\point(\psi)=[T,T,T,F,F,T,F,T,T,T,T]$.
    \item $\intvl(\psi)=[0,2],[5,5],[7,10]$.
      \qed
\end{itemize}
\label{ex5}
\end{example}

\subsection{Algorithm Overview}
Suppose we have $|V|$ freeze variables $x_{*1}, \ldots, x_{*|V|}$ that are ordered in $\varphi$ as follows: $x_{*1} >  x_{*2} > \ldots > x_{*|V|}$, in other words, $\varphi$ is of the form: $\dots x_{*1}.( \dots x_{*2}.(\dots x_{*|V|}.(\dots)))$ (we use $x$ for the naming of the freeze variables instead of $s^k_*$ to avoid any confusion with the subscript $h$ and the signal dimensions superscript $k$).

We will need to consider all possible combinations of environments for all the freeze variables: $\env[x_{*1}:= \sigma_{i_1}^{j_1}, x_{*2}:= \sigma_{i_2}^{j_2}, \dots , x_{*|V|}:= \sigma_{i_{|V|}}^{j_{|V|}}]$ where $i_1 \dots i_V$ are indices referring to timestamps each ranging from $0$ to $|\pi| -1$ and $j_k, 1 \le k \le |V|$ is the index of the signal dimension corresponding to the freeze variable $x_{*k}$. To better understand the idea behind this algorithm, let us consider $\varphi_1$ from example \ref{runningex}, where $x_{*1}=s_{*1}$ and $x_{*2}=s_{*2}$. We consider our first \emph{environment group} which starts with the environment that freezes $s_{*1} \text{ to } \sigma_{0}$ and $s_{*2} \text{ to } \sigma_{0}$, for this first environment, the algorithm calculates the satisfaction relations for every $\psi \in \subtree_{\varphi_0}(s_{*2})$ for all the position indices $i$. Then, the algorithm considers the next environment which freezes $s_{*1} \text{ to } \sigma_{0}$ and $s_{*2} \text{ to } \sigma_{1}$ and calculates the same satisfaction relations for all the position indices $i\ge 1$. Similarly, the next environment would freeze $s_{*1} \text{ to } \sigma_{0}$ and $s_{*2} \text{ to } \sigma_{2}$ and so on $\dots$
Once we reach the environment that freezes $s_{*1} \text{ to } \sigma_{0}$ and $s_{*2} \text{ to } \sigma_{|\pi|-1}$, the algorithm will calculate the satisfaction relation for every $\psi \in \subtree_{\varphi_0}(s_{*1})$ for all the position indices $i$. Then, our next \emph{environment group} would start with the environment that freezes $s_{*1} \text{ to } \sigma_{1}$ and $s_{*2} \text{ to } \sigma_{1}$ and we repeat all the same steps above. The algorithm will go over all the possible environment groups.


For any $x_{*m}>x_{*n}$, if we want to calculate a satisfaction relation for any subformula for a given environment that binds $x_{*m}$ to $\sigma^{j_m}_i$, we only need to consider bindings of $x_{*n}$ to $\sigma^{j_n}_l$ where $i \le l \le |\pi|-1$. That is why, in our previous example of $\varphi_0$, when we moved to the second environment group that freezes $s_{*1}$ to $\sigma_1$, we started by freezing $s_{*2}$ to $\sigma_1$ and not $\sigma_0$.

Here are the main steps of our algorithm:
\begin{itemize}
    \item To be able to get the value of the satisfaction relation $(0, \zeroenv) \models \varphi$, we will need all the values of $(i, \env[\equiv\! \sigma_i]) \models \psi$ for every $\psi \in \myast(\varphi)$ and every $i \in [0, |\pi|-1]$.
    \item And for a given subformula $\varphi_{1} \in \subtree_{\varphi}{(x_{*1})}$, to get the values of $(i, \env[\equiv\! \sigma_i]) \models \varphi_{1}$, we will need all the values of $(j, \env[\equiv\! \sigma_j]) \models \psi$ for every $\psi \in \myast(\varphi_1)$ and every $j \in [0, |\pi|-1]$.
    \item And for a given subformula $\varphi_{2} \in \subtree_\varphi(x_{*2})$, to get the value of $(j, \env[\equiv\! \sigma_j]) \models \varphi_{2}$, we will need all the values of $(k, \env[\equiv\! \sigma_k]) \models \psi$ for every $\psi \in \myast(\varphi_{2})$ and every $k \in [0, |\pi|-1]$.
    \item And so on \dots
\end{itemize}

In general, for a formula with $|V|$ freeze variables, we will go over $|\pi|^{|V|}$ environments for each subformula $\psi \in \subtree_{\varphi}(x_{*|V|})$, and for each $\psi' \in \subtree_{\varphi}(x_{*|V|-1})$, we will be computing $|\pi|^{|V|-1}$ environments $\dots$

The second major contribution of this work is based on the following idea: when trying to calculate $(i, \env) \models \psi$ for any given $\psi$ and any environment $\env$, a naive algorithm would iterate over all the timestamps to calculate the different satisfaction relations. However, our algorithm iterates over the intervals in $\intvl(\psi)$ where in practice the size of $\intvl(\psi)$ (the size of $\intvl(\psi)$ is the number of intervals in $\intvl(\psi)$) is way smaller than the number of timestamps. This will give us the same results in a reduced number of computations. Let us consider example \ref{ex5}, and suppose we want to calculate $(i, \env) \models \Box_{[1,2]} \varphi_0, \forall i$, instead of calculating 10 satisfaction relations (one for each $i$, our algorithm will calculate just 3 (one for each interval in $\intvl(\varphi_0)$, it has just 3 intervals). Also, in some cases, when a subformula $\psi$ is either a signal constraint or of the form $s^k_*.\psi'$, we need to calculate $\point(\psi)$ (the vector $\point(\psi)$ represents $(i, \env) \models \psi, i \ge i'$ for a given $i' \in [0,|\pi|-1]$ and a given $\env$) and not just $\intvl(\psi)$. The nature of the trace $\pi$ (pointwise semantics and discrete timestamps and not a continuous signal) is the main reason why we have to go over $\point(\psi)$ as a first step and not directly calculate $\intvl(\psi)$, in other words, we cannot calculate $\intvl(\psi)$ without calculating $\point(\psi)$ first, for $\psi$ of these forms. For the case of a signal constraint, we try to update a limited number of values in $\point(\psi)$ and not iterate over all values of $i$.

\subsection*{Data Structures}
\begin{itemize}
    \item Timestamps array: Array of size $|\pi|$ which have the timestamps values.
    \item Signal dimension array: For each signal dimension, we use a $|\pi|$ sized array to store the signal dimension values.
    \item $\sorted(\varphi_j)$: Doubly linked list, look section \ref{sorted_section} for more details.
    \item $\point(\varphi_j)$: Array of size $|\pi|$.
    \item $\intvl(\varphi_j)=[\start_j(0),\finish_j(0)],\dots,[\start_j(n),\finish_j(n)] $.
    \item $\start_j$ and $\finish_j$: Arrays of size $n \le |\pi|$ each.
    \item $\flip_j$: Array of size $|\pi|$ for each signal constraint $\varphi_j$.
\end{itemize}

\subsection{$\transform$ Algorithm}
Given input $\point(\varphi_j)$ and integer $i$, $\transform$ computes $\intvl(\varphi_j)$ using the values in $\point(\varphi_j)$ from position $i$.

\subsection{Calculating $\sorted(\varphi')$} \label{sorted_section}
\begin{figure}[h]
    \centering
    \includegraphics[trim=0  0 0 20, scale=0.75]{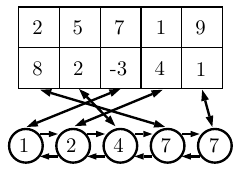}
    \caption{$s^{1}$, $s^{2}$ dimension arrays and $\sorted(\varphi')$}
    \label{fig:list}
\end{figure}

Given any $\varphi'=f_1(\overline{s}) \sim f_2(x_{*1},x_{*2},\dots,x_{*|V|})$ signal constraint, our algorithm needs to calculate $\sorted(\varphi')$. The values in $\sorted(\varphi')$ only depend on the function $f_1$ and the values of $\sigma_i^k, i \in [0,|\pi|-1]$ where $k$ refers to the $k^\text{th}$ signal dimension for every signal dimension called by the function $f_1$.
The algorithm calculates $f_1(\overline{s})$ for the different $i$ values, sorts it and stores it in $\sorted(\varphi')$. Let us consider an example where $f_1= s^{1} + s^{2} - 3$ . The values of $s^{1}$ for the different timestamps are $2,5,7,1,9$ and the values of $s^{2}$ are $8,2,-3,4,1$.
Figure \ref{fig:list} shows the corresponding doubly linked list $\sorted(\varphi')$ and the links between $\sorted(\varphi')$ and the trace values (keep track of original position indices before sorting).

\subsection{Main Algorithm}

\begin{algorithm2e}[htbp]
\small
{\DontPrintSemicolon
\KwIn{$\myast(\varphi)$, $\pi$}
\KwOut{$\intvl(\varphi_1)$}
$\intvl(\varphi') \gets \transform(\point(\varphi'),0)$, $ \forall \varphi'$ signal pred.\;
\For{each sig constraint $\varphi'=f_1(\overline{s}) \sim f_2(x_{*1},\dots,x_{*|V|})$}{$\sorted(\varphi') \gets \text{sorted } f_1(\overline{s}) \text{ values}$} 
Rec-\STLstar(1,0)\;
\If{$|\tsubtree(\varphi)| \not = 1$}{
\For{each subformula $\varphi_j \in \tsubtree(\varphi)$}{$\intvl(\varphi_j) \gets \computesubformula(\varphi_j,0)$}}
\Return $\intvl(\varphi_1)$
 }
\caption{\STLstar Monitor Algorithm}
\label{algo:mainalgo}
\end{algorithm2e}

\begin{algorithm2e}[htbp]{
\small
\KwIn{$k,t$}
\For{$i \gets t$ to $|\pi|-1$}{
    \For{each signal constraint $\varphi_j$ in $\subtree(x_{*k})$}{
        \If{$i=0$}{
        Calculate $\flip_j[0]$, $\point(\varphi_j)$\;
        $\intvl(\varphi_j) \gets \transform(\point(\varphi_j),0)$}
        \lElse{$\intvl(\varphi_j) \gets \updatesignalconstraint(\varphi_j,i)$}}
    \lIf{$k < |V|$}{Rec-\STLstar ($k+1,i$)}
    \For{$j\! \gets\! \subtree_{\varphi}(x_{*k}).\max$ down to $\subtree_{\varphi}(x_{*k}\!).\min$}{\If{$\varphi_j$ is not a sig constraint nor a sig predicate}{$\intvl(\varphi_j) \gets \computesubformula(\varphi_j,i)$}}  

\If{$\intvl(\subtree_{\varphi}(x_{*k}).\roottree)$ starts with $\tau_i$}{$\point(\subtree_{\varphi}(x_{*k}).\parent)[i]  \gets \true$}\lElse{$\point(\subtree_{\varphi}(x_{*k}).\parent)[i]  \gets \false$}
}
 $\intvl(\subtree_{\varphi}(x_{*k}).\parent) \gets \transform(\point(\subtree_{\varphi}(x_{*k}).\parent),t)$
}
\caption{$\recSTLstar(k,t)$}
\label{algo:recSTLstar}
\end{algorithm2e}

The first line of the \STLstar Monitor Algorithm calculates $\intvl(\varphi')$ for any $\varphi'$ signal predicate, the second and third lines calculate $\sorted(\varphi')$ for any $\varphi'$ signal constraint. Line 4 will call the $\recSTLstar$ algorithm using arguments $(1,0)$ to calculate the values of $\intvl(\varphi')$ for every subformula $\varphi'$ in $\subtree_{\varphi}(x_{*k})$ for every freeze variable $x_{*k}$ in $\varphi$ (more details in the next section). The remaining lines in the algorithm (5 to 7) calculate the values of  $\intvl(\varphi')$ for every subformula $\varphi'$ in $\tsubtree({\varphi})$, this is the case when all the subformulae of type $x_{*k}.\psi$ have been computed. Finally, the algorithm returns $\intvl(\varphi_1)$.

Note that if $|\tsubtree(\varphi)| =1$, then the top subtree has only one node of type
$x_{*k}.\psi$, and $\intvl(x_{*k}.\psi)$ has already been computed by $\recSTLstar$ so there is nothing to do.

\subsection{$\recSTLstar(k,t)$}
The $\recSTLstar(k,t)$ function calculates the values of $\intvl(\varphi')$ for all subformulae $\varphi' $ in $\subtree_{\varphi}(x_{*j})$, $k \le j \le |V|$ for the different instantiations of $x_{*k}$ to $\sigma_i^{j_k}$ for $i \in [t,|\pi|-1]$. More simply, the ultimate goal of each call of $\recSTLstar(k,t)$ is to calculate all the $ |\pi|-1 -t$ values of $\point(\subtree_{\varphi}(x_{*k}).\parent)$ (Lines 11-13) and eventually transform it to $\intvl(\subtree_{\varphi}(x_{*k}).\parent)$ (Line 14). Each value in $\point(\subtree_{\varphi}(x_{*k}).\parent)$ represents $(i, \env[x_{*k}:=\sigma_i^{j_k}])  \models \subtree_{\varphi}(x_{*k}).\parent, t \le i \le |\pi|-1$. the main idea behind the $\recSTLstar$ algorithm is the following: in order to calculate $\intvl(\varphi')$ for $\varphi' \in \subtree_{\varphi}(x_{*k})$ at the $i^{\text{th}}$ instantiation, we calculate $\intvl(\varphi')$ for $\varphi' \in \subtree_{\varphi}(x_{*k+1})$ at the instantiations $i'$, $i' \in [i,|\pi|-1]$, that is why, the first time we call $\recSTLstar$, we use arguments (\textbf{1},\textbf{0}) which aim to compute the satisfaction relations of subformulae in  $\subtree_{\varphi}(x_{*\textbf{1}})$ (the first input \textbf{1} refers to the freeze variable $x_{*\textbf{1}}$)  at time points $i \in [\textbf{0},|\pi-1|]$ (the second input \textbf{0} refers to the smallest $i$ value the algorithm considers).

The pre-condition for a function call $\recSTLstar(k,t)$ is that $\intvl(\varphi')$ for any $\varphi'$ signal constraints, belonging to subtrees $\subtree_{\varphi}(x_{*l})$ for $l< k$, have already been computed for $x_{*l}$ instantiated to $\sigma_t^{j_l}$.
Before going through a technical explanation,
let us take an example and suppose our \STLstar formula $\varphi$ is of the form $\varphi=...x_{*1}.(...x_{*2}.(...x_{*3}.(...)))$ where the dots can be any operators and $\varphi$ has 3 freeze variables. The first call $\recSTLstar(1,0)$ will calculate $\intvl(\varphi')$ for any $\varphi'$ signal constraint corresponding to $x_{*1}$ for the instantiation $0$. Then, $\recSTLstar(1,0)$ will call $\recSTLstar(2,0)$ to calculate $\intvl(\varphi')$ for any $\varphi'$ signal constraint corresponding to $x_{*2}$ for the instantiation $0$, and afterwards, $\recSTLstar(3,0)$ is called to calculate $\intvl(\varphi')$ for any signal constraints $\varphi'$ corresponding to $x_{*3}$ for the instantiation $0$. Since $x_{*3}$ is the last freeze variable in $\varphi$,  $\recSTLstar(3,0)$ will continue to calculate $\intvl(\varphi')$ for any subformula $\varphi' \in \subtree_{\varphi}(x_{*3})$ for all the instantiations of $x_{*3}$ to $\sigma_i^{j_3}$ for $i \in [0,|\pi|-1]$. Once that is done, we go back to $\recSTLstar(2,0)$ to calculate $\intvl(\varphi')$ for any $\varphi' \in \subtree_{\varphi}(x_{*2})$ corresponding to $x_{*2}$ instantiated to $\sigma_0^{j_2}$. Then, for $i=1$ in $\recSTLstar(2,0)$, we calculate $\intvl$ for the signal constraints corresponding to $x_{*2}$ for the instantiation $1$ and call $\recSTLstar(3,1)$ which will calculate $\intvl(\varphi')$ for any subformula $\varphi' \in \subtree_{\varphi}(x_{*3})$ again but this time for all instantiations of $x_{*3}$ starting from $1$ and so on \dots

How $\recSTLstar$ is called :
$\recSTLstar(1,0) \to $
$\recSTLstar(2,0) \to$
$\recSTLstar(3,0) \to$
$\recSTLstar(3,1) \ldots \to$
$\recSTLstar(3,|\pi|-1) \to$
$\recSTLstar(2,1) \to$
$\recSTLstar(3,1) \to$
$\recSTLstar(3,2) \ldots \to$
$\recSTLstar(3,|\pi|-1) \ldots \to$
$\recSTLstar(2,|\pi|-1) \to$
$\recSTLstar(3,|\pi|-1)$.

In general, for the instantiation of $x_{*k}$ to $\sigma_i^{j_k}$ (Line 1), we calculate or update the values of $\point(\varphi_j)$ and $\intvl(\varphi_j)$ for every $\varphi_j \in \subtree_{\varphi}(x_{*k})$ signal constraint (Lines 2-6). Then, in a recursive way, the algorithm does the same thing for all the remaining freeze variables $x_{*k'}, k \le k' \le |V|$ instantiated to $\sigma_i^{j_{k'}}$ (Line 7). 
Once the algorithm reaches the final freeze variable $x_{*|V|}$, we already have $\point(\varphi_j)$ and $\intvl(\varphi_j)$ calculated for all $\varphi_j$ signal constraints of the different freeze variables corresponding to all freeze variable instantiated to the corresponding $\sigma_i^{j_{k'}}$, and the algorithm calculates the values of $\intvl(\varphi')$ for every subformula $\varphi' \in \subtree_{\varphi}(x_{*|V|})$ by calling $\computesubformula$ (Lines 8-10) and assigns a value to $\point(\subtree_{\varphi}(x_{*|V|}).\parent)[i]$ (Lines 11-13) ($\point(\subtree_{\varphi}(x_{*|V|}).\parent)[i]$ represents the $i^{\text{th}}$ value in the vector $\point(\subtree_{\varphi}(x_{*|V|}).\parent)$ corresponding to the parent node of $\subtree_{\varphi}(x_{*|V|})$). 
Then, it instantiates $x_{*|V|}$ to the next value $\sigma_{i+1}^{j_{|V|}}$ (next iteration of the \emph{for} loop in Line 1), updates the signal constraints (Line 6), and calculates $\intvl(\varphi')$ for every subformula $\varphi' \in \subtree_{\varphi}(x_{*|V|})$ (Line 10) and adds a new value to $\point(\subtree_{\varphi}(x_{*|V|}).parent)$ and so on until we finish with all the instantiations of $x_{*|V|}$ (Line 1). 
Once that is done, we go back to the previous call of $\recSTLstar(|V|-1,i)$ and calculate the values of $\intvl(\varphi')$ for every subformula $\varphi' \in \subtree_{\varphi}(x_{*|V|-1})$ (and $\point(\varphi')$ in case of a signal constraint). Then, $x_{*|V|-1}$ is instantiated to $i+1$ and we call $\recSTLstar(|V|,i+1)$ and so on.

\subsection{$\computesubformula$ Algorithm} \label{compintvl}
In this section, we show how we compute, for a given environment $\env$, $\intvl(\varphi')$ of subformula $\varphi'$ with boolean or temporal operators. The idea is based on \cite{MalerN04}, we slightly modify it to make it work for pointwise semantics. Suppose we have two traces with different sampling rates. The first one, $\pi_1$, is uniformly sampled of length 100 and the sampling rate is 1 second. And the second one, $\pi_2$, is non-uniformly sampled and it has the following timestamps: $0, 1, 2, 4, 5, 7, 8, 10, 11, 13, 15, 17, 20, 25, 27, 30, 35$ and $40$. And let us consider two signal predicates $\varphi_1 = s^1 \ge 5$ and $\varphi_2=s^2 \le 0$ such that $\intvl(\varphi_1)= [2,10],[20,35]$ and $\intvl(\varphi_2)=[7,15]$, for both traces $\pi_1$ and $\pi_2$.
\subsubsection*{Boolean operators}
For Boolean operators, the computation is straightforward.
We have the following:\\
$\circ$ For the uniformly sampled trace $\pi_1$:
\begin{itemize}
    \item $\intvl(\neg \varphi_1)= [0,1],[11,19],[36,99]$.
    \item $\intvl(\varphi_1 \vee \varphi_2)=[2,15],[20,35]$.
    \item $\intvl(\varphi_1 \wedge \varphi_2)=[7,10]$.
\end{itemize}
$\circ$ For the non-uniformly sampled trace $\pi_2$:
\begin{itemize}
    \item $\intvl(\neg \varphi_1)= [0,1],[11,17],[40,40]$.
    \item $\intvl(\varphi_1 \vee \varphi_2)=[2,15],[20,35]$.
    \item $\intvl(\varphi_1 \wedge \varphi_2)=[7,10]$.
\end{itemize}
For Boolean operators, computing $\intvl(\varphi')$  takes $O(|\intvl(\varphi')|)$.
\subsubsection*{Temporal operators}
To treat temporal operators, we need to define the following \emph{$[a,b]$-back shifting} operation:
\begin{definition}
Let $I = [m, n]$ and $[a, b]$ be intervals and $k$ an index position. The $[a, b]$-back shifting of I, is \[I \ominus [a,b] = [m-b,n-a]\]
We also define the trim of I, $\trim^k(I) $, to be the largest possible interval $[\tau_i,\tau_j],k \le i,j \le |\pi-1|$ included in $I$.\qed
\end{definition}
\noindent \textbf{Note 1:} When we omit the superscript $k$, it means $k=0$.\\
\textbf{Note 2:} For the trim operator, given a $\intvl(\varphi)$ with $|\intvl(\varphi)|$ intervals, if the trace is uniformly sampled (in other words, for a given timestamp, we know the next and previous timestamps in $O(1)$ time), we can calculate $\trim(\intvl(\varphi))$ in $O(|\intvl(\varphi)|)$ time. However, if the trace is not uniformly sampled, calculating $\trim(\intvl(\varphi))$ takes $O(|\intvl(\varphi)|.\text{log}(|\pi|))$ where the $O(\text{log}(|\pi|))$ is paid to find the largest possible interval $[\tau_i,\tau_j],k \le i,j \le |\pi|-1$ included in $I$ for each interval $I$ in $\intvl(\varphi)$ using binary search. Or, we can simply iterate over all the timestamps in $\pi$ to find $\trim(\intvl(\varphi))$ since the intervals in $\intvl(\varphi)$ are ordered. This makes calculating $\trim(\intvl(\varphi))$ takes $O(|\pi|)$. We use the exponential search algorithm, in case of non uniformly sampled trace, to reduce the complexity of calculating $\trim(\intvl(\varphi))$ to the best possible case which is 
$O\Bigl(\min\bigl(|\pi|, |\intvl(\varphi)|\cdot\log(|\pi|)\bigr)\Bigr)$.
\subsubsection*{Eventually operator $\Diamond_{[a,b]}$}
To calculate $\Diamond_{[a,b]} \varphi'$, we just do \\ $\trim(\intvl(\varphi') \ominus [a,b])$. For example, 
\begin{itemize}
    \item For $\pi_1$, $\intvl(\Diamond_{[1,3]} \varphi_1)=[0,9],$ $[17,34]$. 
    \item For $\pi_2$, $\intvl(\Diamond_{[1,3]} \varphi_1)=[0,8],[17,30]$. 
\end{itemize}
\subsubsection*{Until operator $\until_{[a,b]}$}
For $\varphi \until_{[a,b]} \psi$, we will use the same claim used in \cite{MalerN04} for \STL formulae and can be generalized for \STLstar formulae.
\begin{claimnew}
    Let $\varphi=\varphi_1 \vee \varphi_2 \dots \vee \varphi_p$ and $\psi=\psi_1 \vee \psi_2 \dots \vee \psi_q$ be two \STLstar subformula, each written as a union of unitary subformula (with a single interval). Then 
    \[ \varphi \until_{[a,b]} \psi = \bigvee_{i=1}^p \bigvee_{j=1}^q \varphi_i \until_{[a,b]} \psi_j \qed\] 
\end{claimnew} 
\noindent For each interval $I$ in $\varphi$ and $J$ in $\psi$, we do the following: 
$\bigl( (I \cap J) \ominus [a,b] \bigr) \cap I$. Then, we apply the $\trim$ operation to all intervals. \\
For example, let us consider first the uniformly sampled trace $\pi_1$: for $\varphi_1 \until_{[2,4]} \varphi_2$,\\
(a) $[2,10] \cap [7,15]=[7,10]$, $[7,10] \ominus [2,4]=[3,8]$, $[3,8]\cap[2,10]=[3,8]$ and\\
(b) $[20,35] \cap [7,15]= \emptyset$\\
$\Rightarrow \intvl(\varphi_1 \until_{[2,4]} \varphi_2)= [3,8]$.\\
And, for the non-uniformly sampled trace $\pi_2$: we have\\ $\intvl(\varphi_1 \until_{[2,4]} \varphi_2)= [4,8]$.
\begin{itemize}
    \item Uniformly sampled trace: This operation will take \\$O(|\intvl(\varphi_1)|+|\intvl(\varphi_2)|)$.
    \item Non uniformly sampled trace: This operation will take 
    $O(\min(|\pi|, |\intvl(\varphi_1)|+|\intvl(\varphi_2)| .\text{log}|\pi|))$.
\end{itemize}


\subsection{$\updatesignalconstraint$ Algorithm}
Let $\varphi_j$ be a signal constraint, the main goal of this algorithm is to update the values of $\point(\varphi_j)$ and $\intvl(\varphi_j)$ for the instantiation $i+1$ given the values of $\point(\varphi_j)$ and $\intvl(\varphi_j)$ for the instantiation $i$. 

To do so, the algorithm uses $\flip_j$'s to track which values should be updated in $\point(\varphi_j)$ and $\intvl(\varphi_j)$. $\flip_j[i]$ is the position index where $\varphi_j$ $\bigl($in $\sorted(\varphi_j)\bigr)$ switches values from $\true$ to $\false$ or the opposite in the $i^{\text{th}}$ instantiation. 
Here, if we interpret the signal constraint $\varphi_j$ as a function of the frozen values, and since the values are sorted in $\sorted(\varphi_j)$, we can see that $\flip_j[i]$ represents a threshold for when we reach a value in $\sorted(\varphi_j)$ for which $\varphi_j$ is $\true$ (resp. $\false$) for all the next values in $\sorted(\varphi_j)$ and $\false$ (resp. $\true$) for all the previous values.
Given $\flip_j[i]$ and $\flip_j[i+1]$, it updates certain values in $\point(\varphi_j)$ and $\intvl(\varphi_j)$ (values corresponding to position indices between $\flip_j[i]$ and $\flip_j[i+1]-1$ in $\sorted(\varphi)$). Further details in the appendix.

\subsection{Algorithm Correctness} 
The difference between \STLstar monitoring and \STL monitoring is that \STLstar have additional subformulae of type $\psi=x_*.(\dots)$ where we need to consider different freeze environments. In order to obtain the value of $\psi$ at a timestamp $\tau_i$, we need to assign the environment $\env[x_*:= \sigma_i]$ to all subformulae in $\subtree(x_*)$. Once we have the values of $\psi$ for the different timestamps, the remaining steps are similar to an STL monitoring algorithm.

For a given environment $\env[x_*:= \sigma_i]$ and a given $\subtree(x_*)$, a basic algorithm would calculate $(j, \env[x_*:= \sigma_i]) \models \varphi$ for every $j \in [i,|\pi|-1]$, for every $\varphi \in \subtree(x_*)$. Our algorithm uses:\\
$\bullet$ $\updatesignalconstraint$ to update a reduced number of values of satisfaction relations for the signal constraint $\varphi$ from one environment to the next one. It uses the $\flip$ variable to know which indices $j$ for which the value $(j, \env[x_*:= \sigma_i]) \models \varphi$ needs to be updated and which can be skipped. $\point(\varphi)[j]$ and $\intvl(\varphi)$ are updated simultaneously (details in sections \ref{appendix1} and \ref{appendix2}). \\
$\bullet$ $\computesubformula$ and the intervals data structure to accelerate the task of computing $(j, \env[x_*:= \sigma_i]) \models \varphi$ for every $j \in [i,|\pi|-1]$, for every $\varphi \in \subtree(x_*)$ not a signal constraint (details are in section \ref{compintvl} and \cite{MalerN04}).  

Both $\computesubformula$ and $\updatesignalconstraint$ help reduce the number of operations made without changing the output compared to the basic algorithm explained above.

%% file: example.tex
\section{Running Example}
In this section, we will go over the running steps of our algorithm. We will use a formula $\varphi$ similar to $\varphi_1$ from our running example, we slightly modify the signal constraint (node $\varphi_{10}$ in Figure \ref{fig:Tree}) for the sake of simplicity:
    \[\Scale[0.85]{\varphi =\Diamond \biggl(e_1 \wedge s_{*1}.\Bigl(\Diamond \bigl(e_2 \wedge s_{*2}.( \Box_{\ge 2} s \le 0.8\dfrac {s_{*1} + s_{*2}}{2})  \!\bigr)\! \Bigr)\! \biggr)}\]
We consider a uniformly sampled trace with a sampling rate of 1 second. Lines are shown in order as in how the algorithm runs. For non-uniformly sampled traces, the steps are similar with one small difference: when to apply the trim operator when calculating $\intvl(\psi)$ for any subformula $\psi$, this was explained in detail in section \ref{compintvl}. 

In this example, our signal has just one component and it has the following values: $s=(3, 5, 8, 10, 14, 12, 11, 6, 3, 1, 7)$. The algorithm's first step is to calculate $\intvl(\varphi_3)=[3,3]$ and $\intvl(\varphi_7)=[6,6]$ (here, we assume $e_1$ is only $\true$ at $\tau_3$ and $e_2$ at $\tau_6$) and $\sorted(\varphi_{10})=(1, 3, 3, 5, 6, 7, 8, 10, 11, 12, 14)$. 

The algorithm starts with the first environment $\env[s_{*1}:= 3,s_{*2}:= 3]$, computes $\point(\varphi_{10})=[F,F,F,F,F,F,F,F,$\\ $F,T,F]$ (these values are obtained by checking the condition $s \le 0.4 (3 + 3)$ over the different signal values), $\flip_{10}[0]=1$ (this corresponds to the index of the value 3 in $\sorted(\varphi_{10})$, 3 is the lowest $s$ value that does not satisfy $s \le 0.4 (3 + 3)$) and $\intvl(\varphi_{10})=[9,9]$. Then the algorithm calls $\computesubformula$ to calculate $\intvl(\varphi_9)=\emptyset$. After that, it assigns the value $F$ to $\point(\varphi_8)[0]$ which represents the satisfaction relation $(i ,\env[s_{*1}:= 3,s_{*2}:= 3]) \models \varphi_8$.

Then, the algorithm proceeds to the second environment $\env[s_{*1}:= 3,s_{*2}:= 5]$, $\flip_{10}[1]=3$ (now, the lowest $s$ value that does not satisfy $s \le 0.4 (3 + 5)$ is 5). The algorithm should update the value of $\point(\varphi_{10})[0]$ and $\point(\varphi_{10})[8]$ from F to T  corresponding to the signal values with positions 1 and 2 in $\sorted(\varphi_{10})$, however, since we already have the value of $\point(\varphi_8)[0]$, we no longer need to look at or update $\point(\varphi_{10})[0]$ and we can just skip it (more precisely, when freezing $s_{*2}$ to $\sigma_i$, we only update $\point(\varphi_{10})[k], k\ge i$). We obtain  $\point(\varphi_{10})=[-,F,F,F,F,F,F,F,T,T,F]$ and $\intvl(\varphi_{10})=[8,9]$. After that, the algorithm calls $\computesubformula$ to compute $\intvl(\varphi_9)$ then calculates 
$\point(\varphi_8)[1]$.

The algorithm keeps on repeating the previous steps for all the environments $\env[s_{*1}:= 3,s_{*2}:= \sigma_i], 2\le i \le |\pi|-1$ which will result in computing all the values of $\point(\varphi_8)$. With these values, the algorithm is able to compute $\intvl(\varphi_8)$, $\intvl(\varphi_6)$, $\intvl(\varphi_5)$ and $\point(\varphi_4)[0]$ for the environment group that freezes $s_{*1}$ to 3.

Similarly, the algorithm will repeat all the previous steps for the environment group that freezes $s_{*1}$ to $\sigma_i, 1\le i \le |\pi|-1$ in order to compute all $\point(\varphi_4)$ values.

Finally, the algorithm calculates $\intvl(\varphi_2)$ and $\intvl(\varphi_1)$ (Lines 5-7 from algorithm \ref{algo:mainalgo}).

%% file: robust.tex
\section{Quantitative Robustness for \STLstar}
In this section, we define the quantitative semantics for \STLstar via a
\emph{robustness function} $\rho$ which gives a measure of
how well a trace satisfies or violates a given formula. As in the Boolean semantics, to reduce notation clutter, we will ommit $\pi$ and simply write $\rho(\varphi, i, \env)$ instead of $\rho(\varphi, \pi, i, \env)$.

\begin{definition}[Quantitative Semantics]
  \label{definition:Robustness}
   Let $\pi = (\sigma_0, \tau_0), \ldots ,(\sigma_{|\pi|-1}, \tau_{|\pi|-1})$ be a finite
  timed signal of arity $\dm$, $\bowtie_>\in \set{>,\ge}$ and $\bowtie_<\in \set{<,\le}$.
  For a given environment $\env$, and a given index position $0\leq i \leq  |\pi|-1$, the
robustness function valuation $\rho(\varphi, \pi, i, \env) \in \reals$
for a \STLstar formula $\varphi$ is defined as:
\begin{itemize}
  {\setstretch{1.4}
    \item $\!\!\rho(f(\overline{s}) \bowtie_> 0), i, \env) \!=\! f(\overline{\sigma_i})$.
    \item $\!\!\rho(f(\overline{s}) \bowtie_< 0), i, \env) \!=\! - f(\overline{\sigma_i})$.
    \item $\!\!\Scale[0.98]{\rho(f_1(\overline{s}) \bowtie_> f_2(\overline{s_*}) ,i ,\env)=f_1(\overline{\sigma_i}) - f_2(\env(s^{1}_*),...,\env(s^{\dm}_*))}$.
    \item $\!\!\Scale[0.98]{\rho(f_1(\overline{s}) \bowtie_< f_2(\overline{s_*}) ,i ,\env)= f_2(\env(s^{1}_*),...,\env(s^{\dm}_*)) - f_1(\overline{\sigma_i})}$.
  \item $\!\!\rho(\neg \varphi , i, \env) = - \rho(\varphi , i, \env)$.
  \item $\!\!\rho(\varphi_1 \wedge \varphi_2 , i, \env) = \min ( \rho(\varphi_1 , i, \env),\rho(\varphi_2 , i, \env) )$.
  \item $\!\!\rho(\varphi_1 \vee \varphi_2 , i, \env) = \max ( \rho(\varphi_1 , i, \env),\rho(\varphi_2 , i, \env) )$.
  \item $\!\!\rho(\Box_I \varphi , i, \env) =\displaystyle
    \min_{\tau_{j} \in \tau_i+I} (\rho(\varphi , j, \env) )$.
  \item $\!\!\rho(\Diamond_I \varphi , i, \env) =
    \displaystyle \max_{\tau_{j} \in \tau_i+I} (\rho(\varphi , j, \env) )$.

  \item $\!\!\rho(\varphi_1 \until_I \varphi_2 , i, \env) =$
    $\!\Scale[0.88]{\displaystyle\max_{\tau_{j} \in \tau_i+I} \min\! \left(\! \rho(\varphi_2 , j, \env),\min_{\tau_{k} \in [\tau_i,\tau_{j}) }\rho(\varphi_1 , k, \env)\! \right)}$.

  \item $\!\!\rho(s^k_*.\varphi , i, \env)=\rho(\varphi , i, \env[s^k_*:= \sigma_i^k])$.\qed
    }
  \end{itemize}

\end{definition}

\begin{theorem}
  Let $\varphi$ be an \STLstar formula, $i$ a position index, and $\env$ an environment. We have (i)~if $ \rho(\varphi, i,\env) > 0 $ then$ ( i ,\env) \models \varphi$; and
  (ii)~if $ \rho(\varphi, i,\env) < 0 $ then $ ( i ,\env) \not \models \varphi$.
If $\rho(\varphi, i,\env) = 0$, nothing can be concluded. \qed
\end{theorem}

\subsection{\STLstar Quantitative Robustness Monitoring}
\label{subsec:RobustnessMonitoring}
To solve the robustness computation monitoring problem,
we first examine the \emph{decision problem}: is the robustness of a formula for a given trace greater than a given value $r$?

\begin{theorem}[Robustness Decision Problem] \label{robus_theo}
Let $\varphi$ be an \STLstar formula, $i$ a position index, $\env$ an environment and $r$ a real number.  We have (i)~if $( i ,\env) \models  \varphi'$ then $\rho(\varphi, i,\env) \ge r$; and (ii)~if $( i ,\env) \not \models  \varphi'$ then $\rho(\varphi, i,\env) \le r$;
where $\varphi'$ is a logically equivalent syntactic transformation of $\varphi$ obtained as follows: first negation symbols are removed from $\varphi$ by pushing in negation and  reversing any signal constraint or predicate if necessary;
then we replace each of the below subformulae with its corresponding subformula
(where $\bowtie_>\in \set{>,\ge}$ and $\bowtie_<\in \set{<,\le}$).
\begin{itemize}
    {\setstretch{1}
    
    \item $f(\overline{s}) \bowtie_> 0 $ replaced by $f(\overline{s}) \bowtie_> r$.
    \item $f(\overline{s}) \bowtie_< 0 $ replaced by $f(\overline{s}) \bowtie_< - r$.
    \item $f_1(\overline{s}) \bowtie_> f_2(\overline{s_*})$ replaced by $f_1(\overline{s}) \bowtie_> f_2(\overline{s_*}) + r $.
    \item $f_1(\overline{s}) \bowtie_< f_2(\overline{s_*})$ replaced by $f_1(\overline{s}) \bowtie_< f_2(\overline{s_*}) - r $.
    } \qed
  \end{itemize}
\end{theorem} 

Now, given an \STLstar formula and a trace, we find an interval $[a,b]$ for which we are certain that the robustness valuation is within that interval.
\begin{lemma} \label{lemm}
    Given a trace $\pi$, a position index $i \in [0,|\pi|-1]$, an environment $\env$ and a negation-free \STLstar formula $\varphi$,  for any subformula $\psi$ in $\varphi$ we have 
        $\rho(\psi,\pi,i,\env) \in [a,b]$,
    where $a$ and $b$ are the lowest and highest robustness values signal predicates and constraints in $\varphi$ can take over $\pi$. \qed
\end{lemma}

Lemma \ref{lemm} ensures computation of  a conservative range of the robustness value before we run a monitoring algorithm.
We note that any \STLstar formula can be transformed to a negation-free formula as demonstrated in theorem \ref{robus_theo}.

\begin{example}
Consider $\varphi_1$ from example \ref{runningex}. Suppose the highest value of $s$ is 20 and the lowest is 5, then, we have $\rho(\varphi_1,0,\epsilon) \in [-20 + 1.2 \times 5, 1.2 \times 20 - 5]$.~In this example, we give the interval $[a,b]$ just by knowing  $\sup$ and $\inf$ of $s$ since $f_1$ and $f_2$ in the signal constraint $s \in [0.8\dfrac {s_{*1} + s_{*2}}{2},
        1.2\dfrac {s_{*1} + s_{*2}}{2}  ]$ are monotonic. \qed
\end{example}

\noindent\textbf{\textit{Robustness Value Computation}}.
Combining results from Theorem~\ref{robus_theo}, Lemma~\ref{lemm} and the algorithm from Section~\ref{monitor}, we come up with a solution to the quantitative monitoring problem for \STLstar.
Given an \STLstar formula $\varphi$ and a trace $\pi$, we first use Lemma \ref{lemm} to come up with a conservative range $[a,b]$ for $\rho(\varphi,0,\epsilon)$. Next, we do a binary search over the interval $[a,b]$ employing our robustness decision problem monitoring algorithm over different formulae obtained as described in~Theorem \ref{robus_theo}. The different formulae we will use are obtained by picking different $r$ values from Theorem~\ref{robus_theo}. With each new call of the monitoring algorithm, $r$ will be the midpoint of the latest (which is also the smallest so far) conservative range of the robustness value.
We continue the binary search until we reach the desired error range for the robustness value.

%% file: complexity.tex
\section{Running Time}
\subsection{\STLstar Boolean Monitoring Algorithm}
\label{subsec:BooleanMonitoring}
Before giving the complexities of our algorithms, we introduce the following variables:
\begin{itemize}
    \item $|\subtree(x_{*k})|$: \# of sub-formulae in $\subtree(x_{*k})$.
    \item $|V|$: \# of freeze variables in $\varphi$.
    \item $|\intvl(\varphi)|$: maximal number of intervals for any $\varphi_j$ in $\varphi$.
    \item $|\varphi|$: \# of subformulae in $\varphi$.
    \end{itemize}
    We have the following complexities:
 \begin{itemize}
    \item Sorting a signal constraint takes $O\bigl(|\pi|\cdot \log(|\pi|)\bigr)$.
    \item $\transform$ Algorithm: $O(|\pi|)$.
    \item $\updatesignalconstraint$ Algorithm: $O(|\pi|)$.
    \item $\computesubformula$ Algorithm: $O(|\intvl(\varphi)|)$ for a uniformly sampled trace and $O\Bigl(\min\bigl(|\pi|,|\intvl(\varphi)|\cdot \log(|\pi|)\bigr)\Bigr)$ for non uniformly sampled trace.
    \end{itemize}
    
For the Boolean monitoring algorithm, the complexity of the algorithm is the complexity of the recursive algorithm $\recSTLstar(k,t)$. For a given call of $\recSTLstar$ (for a given $k$ and $t$ values), the complexity of Lines 2-6 in $\recSTLstar$ is $O(|\pi|)$ (we assume we have a constant number of signal constraints) and the complexity of Line 8 for loop is $O\bigl(|\varphi|\cdot|\intvl(\varphi)|\bigr)$ for a uniformly sampled trace and $O\Bigl(|\varphi|\cdot \min\bigl(|\pi|,|\intvl(\varphi)|\cdot \log(|\pi|)\bigr)\Bigr)$ for non uniformly sampled trace. And $\recSTLstar(k,t)$ is called $|\pi|^{|V|}$ times. Thus, the complexity of the Boolean monitoring algorithm is:
\begin{itemize}
    \item  $O\Bigl(|\pi|^{|V|} \cdot \max\bigr(|\pi|, \ |\varphi|\! \cdot\! |\intvl(\varphi)|\bigr)\Bigr)$ for uniformly sampled traces.
    \item $\Scale[0.96]{ O \left( |\pi|^{|V|} \cdot \max
      \left( |\pi|,\ \  |\varphi| \!\cdot\! \min\Big(|\pi|,|\intvl(\varphi)|\!\cdot\! \log(|\pi|)\Big) \right) \right)}$ for non-uniform traces.
\end{itemize}

In practice, as we mentioned earlier, we expect $|\intvl(\varphi)|$ to be much smaller compared to trace size $|\pi|$. 

If we drop the intervals idea and  data structure, we obtain a \emph{non-interval} algorithm
inspired by \MTL monitoring algorithms, 
tweaked so that it can be used in the context of \STLstar.
The time complexity of this algorithm is $O\left(c \cdot |\varphi| \cdot |\pi|^{|V|+1}\right)$ (with $c = \lceil a/\Delta \rceil$ where $a$ is the largest constant occurring in the temporal operators in $\varphi$, and $\Delta$ is the smallest difference between two consecutive timestamps, in worst case, $c=|\pi|$). For a given environment and a given subformula, this non-interval algorithm has to compute all satisfaction relations for the different trace points while our algorithm can skip points using the intervals data structure. Additionally, the non-interval algorithm uses recursive formulae for temporal operators \cite{Thati05} which is why we see the new $c$ factor in its complexity (for a given subformula with a timed temporal operator, computing a single satisfaction relation for a given index position and a given environment takes $O(c)$) compared to our algorithm where we avoid it using the intervals data structure. Note here, if one is not careful and does not use the recursive formula for the timed temporal operators, we end up with  complexity $O\left(|\varphi| \cdot |\pi|^{|V|+2}\right)$.

\subsection{\STLstar Quantitative Robustness Computation Algorithm}

\begin{proposition} \label{prop}
    Given an \STLstar formula $\varphi$ and an error value $\epsilon \in \reals_+$, it takes $O(|\pi|^{|V|})$ time to obtain an initial conservative range $[a,b]$ of $\rho(\varphi,0,\env)$ and $n=\lceil\log_2(\frac{b-a}{\epsilon})\rceil$ calls of the monitoring algorithm to obtain a conservative range with a width $\le e$. That range represents our estimation for $\rho(\varphi,0,\env)$. \qed
\end{proposition}
We conclude the following complexities for our \STLstar robustness algorithm is:
\begin{itemize}
    \item  $O\Bigl(n \cdot |\pi|^{|V|} \cdot \max\bigl(|\pi|, |\varphi|\! \cdot\! |\intvl(\varphi)|\bigr)\Bigr)$ for uniformly sampled traces.
    \item{\small $ O \left(n \cdot  |\pi|^{|V|} \cdot \max \Big( |\pi|,\ \  |\varphi|\! \cdot\! \min(|\pi|,|\intvl(\varphi)|\!\cdot\! \log(|\pi|)) \Big) \right)$} for non-uniform traces.
\end{itemize}
We recall the complexity of the algorithm from~\cite{Brim13} is 
$O(|\varphi|\cdot |\pi|^{|V|+2})$. 

%% file: exp.tex
\section{Experiments}

We conducted our experiments on a 64-bit Intel(R) i7-12700H @ 2.30 GHz with 32-GB RAM and we implemented our algorithms using C\texttt{++}. We tested our algorithms on the formulae $\varphi_1 \dots \varphi_4$ from section \ref{expressiveness}:\\
    $\Scale[0.82]{\bullet \text{ }\varphi_1=\Diamond \biggl(e_1 \wedge s_{*1}.\Bigl(\Diamond \bigl(e_2 \wedge s_{*2}. \Box_{[2,T]} s \in [0.8\dfrac {s_{*1} + s_{*2}}{2},
      1.2\dfrac {s_{*1} + s_{*2}}{2}  ] \bigr)\! \Bigr)\!\biggr)}$\\
    $\Scale[0.82]{\bullet \text{ }\varphi_2=\Diamond\biggr( s^1 > 5 \wedge s^2_{*1}.\Diamond \Bigl(s^1 > 10 \wedge s^2_{*2}.\Diamond\bigl( (s^2 > s^2_{*1} + s^2_{*2}) \until s^1 < 5\bigr) \Bigr)\biggr)}$\\
    $\Scale[0.78]{\bullet \text{ }\varphi_3=\Box s_{*1}.\Bigl( \!|s_{*1}\!-\!s| \le \epsilon \text{ } \until \bigl(|s_{*1}\!-\!s| \ge \Delta \wedge s_{*2}.(|s_{*2}\!-\!s| \le \epsilon \text{ } \until
      |s_{*1}\!-\!s| \le \epsilon )\bigr)\!\Bigr)}$\\
    $\Scale[0.83]{\bullet \text{ }\varphi_4=\Box s_{*1}.\Biggl(|s_{*1}-s| \le \epsilon \text{ } \until \biggl(|s_{*1}-s| \ge \Delta \wedge s_{*2}.\Bigl(|s_{*2}-s| \le \epsilon }$ \\[-2mm]
    \hspace*{15mm}$\Scale[0.83]{\until \bigl(|s_{*2}-s| \ge \Delta \wedge \text{ } s_{*3}.(|s_{*3}-s| \le \epsilon \text{ } \until \text{ } |s_{*1}-s| \le \epsilon)\bigr)\Bigr)\biggr)\Biggr)}$\\ 
    We generated the traces using Python.
    Trace noise added by superimposing a noise signal. For each of the 4 formulae above,
     we picked two relevant traces, one of which satisfied the requirement ($\pi_s$) and the other of which violated it ($\pi_v$). For example, for $\varphi_3$, we used traces made of the signal shown in Figure~\ref{fig:signals}.
We ran each of the formulae on different trace sizes:  the same signal over a constant time horizon was sampled with  different sampling rates  to vary the trace size.

\subsection{\STLstar Boolean Monitoring Algorithms}
The traces used in Tables~\ref{tab:results0} are uniformly sampled while the traces in Tables~\ref{tab:results1} are obtained by having random sampling points, in other words, we do not use a fixed sampling rate and the timestamps are randomly selected. We additionally equip our algorithm with an early stoppage condition for formulae starting with $\Box$ or $\Diamond$ (for $\Box \psi$, if $\psi$ is $\false$ once, we return $\false$. Similarly if $\psi$ is $\true$ for $\Diamond \psi$).



 {\begin{table}[ht]
 \smaller
 \scalebox{0.89}{

\begin{tabular}{|c|c|c|c|c|c|c|c|c|c|c|c|c|} 
\hline
  &  & \multicolumn{2}{c|}{$|{\pi}|=500$} & \multicolumn{2}{c|}{$|{\pi}|=1k$} & \multicolumn{2}{c|}{$|{\pi}|=2k$} & \multicolumn{2}{c|}{$|{\pi}|=4k$} & \multicolumn{2}{c|}{$|{\pi}|=10k$}\\
 \hline
\!\!$\varphi$\!\!&\!\!$|\intvl|$\!\!& $\pi_s$ & $\pi_v$& $\pi_s$ & $\pi_v$& $\pi_s$ & $\pi_v$& $\pi_s$ & $\pi_v$& $\pi_s$ & $\pi_v$\\
 \hline 
 \!\!$\varphi_1$\!\!& 5 &\!0.04\!&\!0.36\!&\!0.22\!&\!1.47\!&\!0.96\!&\!5.88\!& 3.86\!&\!24\!&\!24.2\!&\!151\!\\
 \hline 
 \!\!$\varphi_2$\!\!& 6 &\!0.04\!&\!0.37\!&\!0.18\!&\!1.52\!&\!0.72\!&\!6.06\!&\!2.88\!&\!24\!&\!18.2\!&\!152\!\\
 \hline 
 \!\!$\varphi_3$\!\!& 11 &\!0.45\!&\!0.02\!&\!1.84\!&\!0.11\!&\!7.31\!&\!0.44\!&\!29\!&\!1.76\!&\!183\!&\!11\!\\
 \hline 
 \!\!$\varphi_4$\!\!& 22  &\!157\!&\!14.2\!&\!\!22m\!\!&\!\!2m\!\!&\!\!178m\!\!&\!\!17m\!\!& - & - & - & -\\
  \hline 
\end{tabular}}
\caption{Running times in seconds (or minutes, m) of \STLstar Boolean monitoring algorithm over uniformly sampled traces. Trace $\pi_s$ satisfies the formulae and $\pi_v$ violates it.}
\label{tab:results0}
\end{table}}
{\begin{table}[ht]
 \smaller
\centering
\scalebox{0.89}{

\begin{tabular}{|c|c|c|c|c|c|c|c|c|c|c|c|c|} 
\hline
  &  & \multicolumn{2}{c|}{$|{\pi}|=500$} & \multicolumn{2}{c|}{$|{\pi}|=1k$} & \multicolumn{2}{c|}{$|{\pi}|=2k$} & \multicolumn{2}{c|}{$|{\pi}|=4k$} & \multicolumn{2}{c|}{$|{\pi}|=10k$}\\
 \hline
\!\!$\varphi$\!\!&\!\!$|\intvl|$\!\!& $\pi_s$ & $\pi_v$& $\pi_s$ & $\pi_v$& $\pi_s$ & $\pi_v$& $\pi_s$ & $\pi_v$& $\pi_s$ & $\pi_v$\\
 \hline 
 \!\!$\varphi_1$\!\!& 5 &\!0.04\!&\!0.36\!&\!0.22\!&\!1.46\!&\!0.96\!&\!5.90\!&\!3.85\!&\!24\!&\!24.2\!&\!151\!\\
 \hline 
 \!\!$\varphi_2$\!\!& 6 &\!0.04\!&\!0.38\!&\!0.19\!&\!1.53\!&\!0.73\!&\!6.10\!&\!2.89\!&\!24\!&\!18.3\!&\!153\!\\
 \hline 
 \!\!$\varphi_3$\!\!& 11 &\!0.45\!&\!0.02\!&\!1.87\!&\!0.12\!&\!7.45\!&\!0.47\!&\!30\!&\!1.78\!&\!184\!&\!11.1\!\\
 \hline 
 \!\!$\varphi_4$\!\!& 22  &\!158\!&\!14.3\!&\!23m\!&\!2m\!&\!184m\!&\!18m\!& - & - & - & -\\
  \hline 
\end{tabular}}
  \caption{Running times of \STLstar Boolean monitoring algorithm over non-uniformly sampled traces.}
\label{tab:results1}
\end{table}}

We also implemented a Boolean version of the \STLstar robustness monitoring algorithm from~\cite{Brim13} and equipped it with the same early stoppage condition to make the comparison fair. We report the running times in Table \ref{tab:results4}. Note that that our formulae do not have timed until operators, that is why the complexity of the non-interval algorithm from~\cite{Brim13} in this case is $O\left(|\varphi| \cdot |\pi|^{|V|+1}\right)$. In the general case, the complexity would be $O\left(|\varphi| \cdot |\pi|^{|V|+2}\right)$.

 {\begin{table}[ht]
 \smaller
 \scalebox{0.93}{

\begin{tabular}{|c|c|c|c|c|c|c|c|c|c|c|c|} 
\hline
  &  \multicolumn{2}{c|}{$|{\pi}|=500$} & \multicolumn{2}{c|}{$|{\pi}|=1k$} & \multicolumn{2}{c|}{$|{\pi}|=2k$} & \multicolumn{2}{c|}{$|{\pi}|=4k$} & \multicolumn{2}{c|}{$|{\pi}|=10k$}\\
 \hline
\!\!$\varphi$\!\! & $\pi_s$ & $\pi_v$& $\pi_s$ & $\pi_v$& $\pi_s$ & $\pi_v$& $\pi_s$ & $\pi_v$& $\pi_s$ & $\pi_v$\\
 \hline 
 \!\!$\varphi_1$\!\!&\!0.21\!&\!1.84\!&\!2.07\!&\!14.8\!&\!16.9\!&\!119\!&\!121\!&\!16m\!&\!37m\!&\!262m\!\\
 \hline 
 \!\!$\varphi_2$\!\!&\!0.18\!&\!1.80\!&\!1.52\!&\!14.4\!&\!14.7\!&\!116\!&\!114\!&\!16m\!&\!31m\!&\!255m\!\\
 \hline 
 \!\!$\varphi_3$\!\!&\!1.82\!&\!0.11\!&\!14.6\!&\!0.88\!&\!118\!&\!7.12\!&\!16m\!&\!56\!&\!255m\!&\!16m\!\\
 \hline 
 \!\!$\varphi_4$\!\!&\!462\!&\!41.7\!&\!124m\!&\!11m\!& -& - & - & - & - & -\\
  \hline 
\end{tabular}}
\caption{Running times in seconds (or minutes, m) of non-interval \STLstar Boolean monitoring algorithm.}
\label{tab:results4}
\end{table}}
Our experimental results show that our algorithm outperforms the non-interval algorithm. We can see that, in practice, the number of intervals $|\intvl(\varphi_i)|$ is much smaller compared to $|\pi|$. In addition, $|\intvl(\varphi_i)|$ is independent of the trace size which is expected since we use same signals with the same time horizon, with different sampling rates. We also notice that the early stoppage condition helps reduce the running times significantly.

The formulae $\varphi_1, \varphi_2$ and $\varphi_3$ have 2 freeze variables and we can see, for the cases where there is no early stoppage, an almost quadratic running time in proportion with the trace size, however, when we look at the previous complexity analysis, it indicates a cubic dependence. This can be explained as follows: $\updatesignalconstraint$ often  does not require $O(|\pi|)$ running time in practice: from one instantiation to the next one, only few values, and not all $|\pi|$ values, in $\point(\psi)$ (where $\psi$ is the signal constraint) will need updates while the majority of values will remain the same.
This is due to the fact that signals in real-world systems are continuous, and in our case for pointwise semantics, from one timestamp to the next one, we do not expect large trace value change. Thus, from one environment to the next, the values of the $f_2$ function in the signal constraints do not have sudden shifts (for example, if we had non-continuous $f_2$ functions in our signal constraints, we would expect a higher number of values needs to be updated).
Hence, we expect $\updatesignalconstraint$ to run in $O(\log(|\pi|))$, the time needed to find the new value of $\flip_j$ using $\sorted(\varphi_j)$. 
With that assumption, the complexity can be simplified to
\begin{itemize}
    \item  $O\Bigl(|\pi|^{|V|} \cdot \max\bigr(\log(|\pi|), |\varphi| . |\intvl(\varphi)|\bigr)\Bigr)$ for a uniformly sampled trace.
    \item $ O \Biggl( |\pi|^{|V|} \cdot \max \left(\begin{array}{l} \log(|\pi|),\\ |\varphi| \cdot \min(|\pi|,|\intvl(\varphi)|\cdot \log(|\pi|)) \end{array}\right) \Biggr)$ or simply $ O\Bigl( |\pi|^{|V|} \cdot |\varphi| \cdot \min\bigl(|\pi|,|\intvl(\varphi)|\cdot \log(|\pi|)\bigr) \Bigr)$ for a non-uniformly sampled trace.
\end{itemize}

For $\varphi_4$ which has 3 freeze variables, our algorithm starts running slow once the trace gets larger, which is to be expected from our complexity analysis.

\subsection{\STLstar Quantitative Robustness Computation Algorithms}
For computing robustness, we implemented two algorithms:
(i)~a non-interval \STLstar robustness monitoring algorithm from~\cite{Brim13} (results in Table \ref{tab:results3}); and
(ii)~the interval  \STLstar robustness monitoring algorithm from Subsection~\ref{subsec:RobustnessMonitoring} (results in Table~\ref{tab:results2}).
The experiments were run on uniformly sampled traces; times for non-uniformly sampled traces are expected to
be similar  based on Table \ref{tab:results0} and Table \ref{tab:results1} results.


\begin{table}[ht]
 \hspace*{-2mm} {
      \scalebox{0.82}{
        \begin{tabular}{|c|c|c|c|c|c|} 
          \hline
          $\varphi$ &   $|{\pi}|=500$   &   $|{\pi}|=1k$   &   $|{\pi}|=2k$   &   $|{\pi}|=4k$   &   $|{\pi}|=10k$  \\
          \hline
          $\varphi_1$  &  1.91 & 15.31 &  123 &  17m & 266m\\
          \hline 
          $\varphi_2$  &  1.83 & 14.65 &   118 &  16m & 252m\\
          \hline 
          $\varphi_3$  &   1.94 & 15.52 &  124 &  17m & 267m\\
          \hline 
          $\varphi_4$  &  475 & 126m &  -  &  - &  -\\
          \hline 
        \end{tabular}}
      \caption{Running times in seconds   (or minutes, m) of non-interval \STLstar robustness monitoring algorithm.}
      \label{tab:results3}
      }
    \end{table}
    

    \noindent\textbf{Non-interval \STLstar robustness monitoring algorithm~\cite{Brim13}}.
    This algorithm has the following complexity $O\left(|\varphi| \cdot |\pi|^{|V|+2}\right)$. 
    It uses the $\max/\min$ filter from \cite{lemire06} to reduce the complexity to $O\left(|\varphi| \cdot |\pi|^{|V|+1}\right)$ for formulae without timed until operator. We recall that in our experiments, all formulae do not have timed until. We note that this algorithm cannot have the early stoppage condition for $\Box$ and $\Diamond$ since this option can only be applied for the Boolean case.
    The experiment results are in
    Table~\ref{tab:results3}.

    \smallskip 
    Our interval \STLstar robustness monitoring algorithm from  Subsection~\ref{subsec:RobustnessMonitoring}
        stops the binary search once it reaches a conservative range  $e \le 0.1$ for the robustness.  In Table \ref{tab:results2}, we also report the relative error of the estimated robustness value compared to the exact value.
\begin{table}[ht]
  \vspace*{2mm}
  \hspace*{-2mm}{
\scalebox{0.8}{
\begin{tabular}{|c|c|c|c|c|c|c|c|c|c|} 
\hline
  \!\!$\varphi$\!\!&\!\!{\small$|\intvl|$}\!\!&\!\!$n$\!\!&\!\!{\small i.c.r.w.}\!\!  &   \!\!r.e\!\!   &\!\!{\small$| {\pi}|=500$}\!\!&\!\!{\small$|{\pi}|=1k$}\!\!& \!\!{\small$|{\pi}|=2k$}\!\!&\!\!{\small$|{\pi}|=4k$}\!\!&\!\!{\small$|{\pi}|=10k$}\!\!\\
\hline
 \!$\varphi_1$\!\!& 8 & \!\!9\!\!& 49 &\!1\%\!\!&  2.16 & 9.02 & 38.1 & 149 & 15m\\
 \hline 
 \!$\varphi_2$\!\!& 8 &\!\!10\!\!& 83 &\!$\le$\,1\%\!\!&  2.31& 9.38 & 39.5 & 154 & 16m\\
 \hline 
 \!$\varphi_3$\!\!& 13 &\!\!9\!\!& 43 &\!2\%\!\!&  2.47 & 10.3 & 41 & 166 & 17m\\
 \hline 
 \!$\varphi_4$\!\!& 25 &\!\!9\!\!& 173 &\!$\le$\,1\%\!\!&   15m  &\!126m\!& - & -& -\\
  \hline 
\multicolumn{10}{l}{\rule{0pt}{3ex}   
\normalsize	   \hspace*{-1mm}$n$: number of times the \STLstar Boolean algorithm is called.} \\
\multicolumn{10}{l}{   
\normalsize i.c.r.w: initial conservative range width, $b-a$ in proposition \ref{prop}.}  \\
\multicolumn{10}{l}{   
\normalsize r.e: estimated robustness value relative error.} \\
\end{tabular}}
  \caption{Running times in seconds  (or minutes, m) of interval \STLstar robustness monitoring algorithm.}
  \label{tab:results2}
  \vspace*{1mm}
}
\end{table}
The obtained $n$ values in Table~\ref{tab:results2} confirm our analysis in proposition \ref{prop}. We can also see that the values for $\intvl(\varphi)$ are slightly larger compared to the previous experiments, this can be explained by the fact that we are running the Boolean monitoring algorithm over slightly different formulae and not the original formulae $\varphi_1 \dots \varphi_4$. The new modified formulae give us in some cases lower $\intvl(\varphi)$ value and in other cases higher value.
The running times conform with our complexity analysis and show that, in most cases, the accelerated \STLstar robustness computation algorithm scales better than the non-interval one.

%% file: conclusion.tex
\section{Conclusion}
In this work we presented an acceleration heuristic using intervals for monitoring \STLstar specifications. We showed that with this heuristic, monitoring for \STLstar specifications with two nested freeze variables remains tractable for Boolean monitoring as well as for robustness monitoring; and somewhat tractable for three nested freeze variables.
We posit engineering properties of interest that can be expressed in \STLstar can be expressed in the \STLstar  subset of two, and occasionally three, nested freeze variables.
Ours is the \emph{first} work which presents implemented Boolean  and robustness monitoring algorithms for formulae with nested freeze quantifiers.
For the  robustness value computation, we first presented algorithms for the corresponding decision problem using the acceleration heuristic, and then computed the robustness value using binary search.
A notable feature of using the decision problem procedure for the robustness value computation is that it allows early stoppage for $\Box$ and $\Diamond$ operators; such early stoppage is not possible in a direct robustness value computation which does not use the decision problem algorithm.

One of the main applications of temporal logic robustness is in the test-generation setting where black-box optimizers are used to search for an input such that the corresponding
system output robustness value is negative, falsifying the logical specification~\cite{AkazakiH15,AnnpureddyLFS11,BarbotBDDKY20,DokhanchiYHF17,RamezaniDFA21,Waga20,ZhangAH20}. In such a setting, one could conceivably stop the
robustness  binary searches earlier -- and thus gain even more in terms of time --  if the robustness value range is positive and greater than the robustness values seen for previous inputs, as we aim to search for an input that drives the robustness value lower than those seen so far. How this would impact the falsification process depends on the optimizer used and how it uses the actual robustness values. We plan to investigate this line of research in follow-up work.

%% file: appendix.tex
\section*{Appendix}
\begin{definition}[Syntax Tree]
  \label{def:AST}
  Given an \STLstar formula $\varphi$, the associated abstract syntax tree $\myast(\varphi)$
  is defined as follows.
  \begin{itemize}
  \item The nodes of the syntax tree are $\sub(\varphi)$.
  \item The root node is $\varphi$.
  \item The edges in the tree are defined by the operator structure:
    \begin{itemize}


    \item If $s^k_{*}. \psi \in \sub(\varphi)$, then $s^k_{*}.\psi$ has the child $ \psi$.

    \item If $\op \psi \in \sub(\varphi)$, for $\op\in \set{\neg, \Box_I, \Diamond_I}$,

      then $\op \psi$ has the child $ \psi$.


    \item If $\psi_1 \op \psi_2 \in \sub(\varphi)$, for $\op\in \set{\wedge, \vee,\to, \until_I}$,
      then $\psi_1 \op \psi_2 $
      has the two children $\psi_1, \psi_2$. \qed


      
    \end{itemize}
  \end{itemize}

  \end{definition}

\subsection{$\updatesignalconstraint$ Algorithm} \label{appendix1}
Let $\varphi_j$ be a signal constraint, the main goal of this algorithm is to calculate the satisfaction relation $(i+1 ,\env) \models \varphi_j$ given the satisfaction relation $(i,\env) \models \varphi_j$. 

Given $\flip_j[i]$ and $\flip_j[i+1]$, it updates certain values in $\point(\varphi_j), \start_j$ and $\finish_j$ (values corresponding to position indices between $\flip_j[i]$ and $\flip_j[i+1]-1$ in $\sorted(\varphi)$) by calling $\subupdate$ (Lines 1-4).

Finally, the algorithm either sorts the values in $\start_j$ and $\finish_j$ to get $\intvl(\varphi_j)$ (Lines 5-7) (since the values in $\intvl(\varphi_j)$ are initially from the previous instantiation, it could be that the first interval in $\intvl(\varphi_j)$ starts with $\tau_{i-1}$, we use the operation in line 7 to make sure that it starts with $\tau_{i'}$ where $i' \ge i$) or just calculates $\start_j$ and $\finish_j$ from scratch using $\point(\varphi_j)$ (Lines 8-10), depending on which operation is estimated to be faster. In fact, in some cases, $\start_j$ and $\finish_j$ can be too long (we use the condition in line 5) and it is better to remove all the values from $\start_j$ and $\finish_j$, and iterate over $\point(\varphi_j)$ to get the new values sorted (Line 6 takes $O(size(\start_j)\cdot\log(size(\start_j)))$ while Line 10 takes $O(|\pi|)$).  

\begin{algorithm2e}[htbp]
\small
\DontPrintSemicolon
\KwIn{$\intvl(\varphi_j)$ in $i^{\text{th}}-1$ instantiation, $i$}
\KwOut{$\intvl(\varphi_j)$ in $i^{\text{th}}$ instantiation}
Calculate $\flip_j[i]$\;
\For{each position index $l$ between $\flip_j[i-1]\text{ and }\flip_j[i]-1 $ in $\sorted(\varphi_j)$}{
\If{$l\ge i$}{
$\start_j,\finish_j,\point(\varphi_j) \gets \subupdate(\tau_l,\start_j,\finish_j)$}}
\If{size$(\start_j).\text{log(size}(\start_j))<|\pi|$}{sort $\start_j$ and $\finish_j$\;
$\intvl(\varphi_j) \gets \intvl(\varphi_j) \cap [\tau_{i}, \tau_{|\pi-1|}]$
}
\Else{empty $\start_j$ and $\finish_j$ \;
$\intvl(\varphi_j) \gets \transform(\point(\varphi_j),i)$
}
\Return $\intvl(\varphi_j)$
\caption{$\updatesignalconstraint$ }
\label{algo:sig_const}
\end{algorithm2e}

\subsection{$\subupdate$ Algorithm} \label{appendix2}
Given $\point(\varphi_j),\start_j,\finish_j$ and a position index $l$, the goal of this algorithm is to, first (Line 6 or 12), update the value $\point(\varphi_j)[l]$ corresponding to the satisfaction relation \\ $(l,\env[x_{*k} := \sigma_{i}^{j_k}]) \models \varphi_j$ ($i$ is the current value when \\ $\updatesignalconstraint$ calls $\subupdate$). And second (Lines 1-5 or 7-11), 
make the necessary changes to $\start_j$ and $\finish_j$ so that $\intvl(\varphi_j)$ is also updated and keeping track of the changes happening to $\point(\varphi_j)$.

We use two  basic operations on $\start_j$ and $\finish_j$, $\add$ and $\remove$. For $\remove$, it is a ``lazy" remove: instead of removing an element right away when we call $\remove$, we only do that once we call sort on the array.



Let us consider the example $\intvl(\varphi_1)=[2,10],$ $[20,35]$ (in other words $\start_1=[2,20]$ and $\finish_1[10,35]$) and suppose the value $\point(\varphi_1)[8]$ need to change from $\true$ to $\false$. Then the algorithm will change the value $\point(\varphi_1)[8]$ to $\false$ (Line 6). The conditions in Lines 2 and 4 are satisfied so the
And the algorithm will add the value 9 to $\start_1$ and the value 7 to $\finish_1$ end we end up with  $\start_1=[2,20,9]$ and $\finish_1[10,35,7]$. and once we sort $\finish_1$ and $\start_1$ (this is done in the $\updatesignalconstraint$ algorithm), we get $\intvl(\varphi_1)= [2,7],[9,10],[20,35]$.


\begin{algorithm2e}[h]
\small
\DontPrintSemicolon
\KwIn{$\tau_l,\start_j,\finish_j$}
\KwOut{$\start_j,\finish_j,\point(\varphi_j)$}
\If{$\point(\varphi_j)[l]=\true$}{
\lIf{$\point(\varphi_j)[l+1]=\true$}{$\start_j.\add(\tau_{l+1})$}
\lElse{$\finish_j.\remove(\tau_l)$}
\lIf{$\point(\varphi_j)[l-1]=\true$}{$\finish_j.\add(\tau_{l-1})$}
\lElse{$\start_j.\remove(\tau_l)$}
$\point(\varphi_j)[l] \gets \false$
}
\If{$\point(\varphi_j)[l]=\false$}{
\lIf{$\point(\varphi_j)[l+1]=\false$}{$\finish_j.\add(\tau_{l})$}
\lElse{$\start_j.\remove(\tau_{l+1})$}
\lIf{$\point(\varphi_j)[l-1]=\false$}{$\start_j.\add(\tau_{l})$}
\lElse{$\finish_j.\remove(\tau_{l-1})$}
$\point(\varphi_j)[l] \gets \true$
}
\Return $\start_j,\finish_j,\point(\varphi_j)$
\caption{$\subupdate$}
\label{algo:update}
\end{algorithm2e}

\vspace{5mm}
\textbf{Theorem \ref{robus_theo} proof}: We show that any \STLstar formula $\varphi$ can be transformed into another \STLstar formula that has no negation operators. First, we push all the negations in $\varphi$ to the signal predicates and constraints as follows:
\begin{itemize}
    \item $\neg (\varphi_1 \wedge \varphi_2) \equiv \neg \varphi_1 \vee \neg \varphi_2$.
    \item $\neg (\varphi_1 \vee \varphi_2) \equiv \neg \varphi_1 \wedge \neg \varphi_2$.
    \item $\neg \Box_I \varphi \equiv \Diamond_I \neg \varphi$.
    \item $\neg \Diamond_I \varphi \equiv \Box_I \neg \varphi$.
    \item $\neg (\varphi_1 \until_I \varphi_2) \equiv \Box_I \neg \varphi_2 \vee \neg \varphi_2 \until_I (\neg \varphi_1 \wedge \neg \varphi_2)$.
\end{itemize}
Then, for the signal predicates and constraints, we reverse the comparison operator and remove the negation.

Suppose $\varphi$ is a subformula of the form $f_1(\overline{s}) \sim f_2(\overline{s_*})$
$\bullet \text{ Case }(i,\env) \models f_1(\overline{s}) > f_2(\overline{s_*}) +r\\ 
\hspace*{6mm} \text{ iff }f_1(\overline{\sigma_i}) > f_2(\env(s^{1}_*),...,\env(s^{\dm}_*)) + r \\
\hspace*{6mm} \text{ iff }f_1(\overline{\sigma_i}) - f_2(\env(s^{1}_*),...,\env(s^{\dm}_*)) >  r \\
\hspace*{6mm} \text{ iff }\rho(\varphi,i,\env) > r \\
\hspace*{6mm} \text{ then }\rho(\varphi,i,\env) \ge r  
$\\
$
\bullet \text{ Case }(i,\env) \models f_1(\overline{s}) \ge f_2(\overline{s_*}) +r 
\\ \hspace*{6mm} \text{ iff }f_1(\overline{\sigma_i}) \ge f_2(\env(s^{1}_*),...,\env(s^{\dm}_*)) + r \\
\hspace*{6mm} \text{ iff }f_1(\overline{\sigma_i}) - f_2(\env(s^{1}_*),...,\env(s^{\dm}_*)) \ge  r \\
\hspace*{6mm} \text{ iff }\rho(\varphi,i,\env) \ge r
$\\
$
\bullet \text{ Case }(i,\env) \models f_1(\overline{s}) < f_2(\overline{s_*}) -r\\ 
\hspace*{6mm} \text{ iff }f_1(\overline{\sigma_i}) < f_2(\env(s^{1}_*),...,\env(s^{\dm}_*)) - r \\
\hspace*{6mm} \text{ iff }f_2(\env(s^{1}_*),...,\env(s^{\dm}_*)) - f_1(\overline{\sigma_i}) >  r \\
\hspace*{6mm} \text{ iff }\rho(\varphi,i,\env) > r \\
\hspace*{6mm} \text{ then }\rho(\varphi,i,\env) \ge r  
$\\
$
\bullet \text{ Case }(i,\env) \models f_1(\overline{s}) \le f_2(\overline{s_*}) -r \\
\hspace*{6mm} \text{ iff }f_1(\overline{\sigma_i}) \le f_2(\env(s^{1}_*),...,\env(s^{\dm}_*)) - r \\
\hspace*{6mm} \text{ iff }f_2(\env(s^{1}_*),...,\env(s^{\dm}_*)) - f_1(\overline{\sigma_i}) \ge  r \\
\hspace*{6mm} \text{ iff }\rho(\varphi,i,\env) \ge r  
$\\
$
\bullet \text{ Case }(i,\env) \not \models f_1(\overline{s}) > f_2(\overline{s_*}) +r \\
\hspace*{6mm} \text{ iff }f_1(\overline{\sigma_i}) \le f_2(\env(s^{1}_*),...,\env(s^{\dm}_*)) + r \\
\hspace*{6mm} \text{ iff }f_1(\overline{\sigma_i}) - f_2(\env(s^{1}_*),...,\env(s^{\dm}_*)) \le  r \\
\hspace*{6mm} \text{ iff }\rho(\varphi,i,\env) \le r  
$\\
$
\bullet \text{ Case }(i,\env) \not \models f_1(\overline{s}) \ge f_2(\overline{s_*}) +r \\
\hspace*{6mm} \text{ iff }f_1(\overline{\sigma_i}) < f_2(\env(s^{1}_*),...,\env(s^{\dm}_*)) + r \\
\hspace*{6mm} \text{ iff }f_1(\overline{\sigma_i}) - f_2(\env(s^{1}_*),...,\env(s^{\dm}_*)) <  r \\
\hspace*{6mm} \text{ iff }\rho(\varphi,i,\env) < r \\
\hspace*{6mm} \text{ then }\rho(\varphi,i,\env) \le r 
$\\
$
\bullet \text{ Case }(i,\env) \not \models f_1(\overline{s}) < f_2(\overline{s_*}) -r \\
\hspace*{6mm} \text{ iff }f_1(\overline{\sigma_i}) \ge f_2(\env(s^{1}_*),...,\env(s^{\dm}_*)) - r \\
\hspace*{6mm} \text{ iff }f_2(\env(s^{1}_*),...,\env(s^{\dm}_*)) - f_1(\overline{\sigma_i}) \le  r \\
\hspace*{6mm} \text{ iff }\rho(\varphi,i,\env) \le r  
$\\
$
\bullet \text{ Case }(i,\env) \not \models f_1(\overline{s}) \le f_2(\overline{s_*}) -r\\ 
\hspace*{6mm} \text{ iff }f_1(\overline{\sigma_i}) > f_2(\env(s^{1}_*),...,\env(s^{\dm}_*)) - r \\
\hspace*{6mm} \text{ iff }f_2(\env(s^{1}_*),...,\env(s^{\dm}_*)) - f_1(\overline{\sigma_i}) <  r \\
\hspace*{6mm} \text{ iff }\rho(\varphi,i,\env) < r  \\
\hspace*{6mm} \text{ then }\rho(\varphi,i,\env) \le r  
$\\
Similarly if $\varphi$ is a subformula of the form $f(\overline{s}) \sim 0$.\\ Now, let $\varphi_1$ and $\varphi_2$ two \STLstar formulae and suppose that    
(i)~if $( i ,\env) \models  \varphi'_1$ then $\rho(\varphi_1, i,\env) \ge r$; (ii)~if $( i ,\env) \not \models  \varphi'_1$ then $\rho(\varphi_1, i,\env) \le r$; (iii)~if $( i ,\env) \models  \varphi'_2$ then $\rho(\varphi_2, i,\env) \ge r$; and (iv)~if $( i ,\env) \not \models  \varphi'_2$ then $\rho(\varphi_2, i,\env) \le r$.\\
$
\bullet \text{ Case }(i,\env)  \models \varphi'_1 \wedge \varphi'_2\\
\hspace*{6mm} \text{ iff }(i,\env) \models \varphi'_1 \text{ and } (i,\env)  \models \varphi'_2 \\
\hspace*{6mm} \text{ iff }\rho(\varphi_1, i,\env) \ge r \text{ and } \rho(\varphi_2, i,\env) \ge r \\
\hspace*{6mm} \text{ iff }\min(\rho(\varphi_1 , i,\env),\rho(\varphi_2 , i,\env)) \ge r  \\
\hspace*{6mm} \text{ iff }\rho(\varphi_1 \wedge \varphi_2 , i,\env) \ge r  $  \\
$\bullet \text{ Case }(i,\env) \not \models \varphi'_1 \wedge \varphi'_2\\
\hspace*{6mm} \text{ iff }(i,\env) \not \models \varphi'_1 \text{ or } (i,\env)  \not \models \varphi'_2 \\
\hspace*{6mm} \text{ iff }\rho(\varphi_1, i,\env) \le r \text{ or } \rho(\varphi_2, i,\env) \le r \\
\hspace*{6mm} \text{ iff }\min(\rho(\varphi_1 , i,\env),\rho(\varphi_2 , i,\env)) \le r  \\
\hspace*{6mm} \text{ iff }\rho(\varphi_1 \wedge \varphi_2 , i,\env) \le r  
$\\
$
\bullet \text{ Case }(i,\env)  \models \varphi'_1 \vee \varphi'_2\\
\hspace*{6mm} \text{ iff }(i,\env) \models \varphi'_1 \text{ or } (i,\env)  \models \varphi'_2 \\
\hspace*{6mm} \text{ iff }\rho(\varphi_1, i,\env) \ge r \text{ or } \rho(\varphi_2, i,\env) \ge r \\
\hspace*{6mm} \text{ iff }\max(\rho(\varphi_1 , i,\env),\rho(\varphi_2 , i,\env)) \ge r  \\
\hspace*{6mm} \text{ iff }\rho(\varphi_1 \vee \varphi_2 , i,\env) \ge r  
$\\
$
\bullet \text{ Case }(i,\env) \not \models \varphi'_1 \vee \varphi'_2\\
\hspace*{6mm} \text{ iff }(i,\env) \not \models \varphi'_1 \text{ and } (i,\env)  \not \models \varphi'_2 \\
\hspace*{6mm} \text{ iff }\rho(\varphi_1, i,\env) \le r \text{ and } \rho(\varphi_2, i,\env) \le r \\
\hspace*{6mm} \text{ iff }\max(\rho(\varphi_1 , i,\env),\rho(\varphi_2 , i,\env)) \le r  \\
\hspace*{6mm} \text{ iff }\rho(\varphi_1 \vee \varphi_2 , i,\env) \le r  
$\\
$
\bullet \text{ Case }(i,\env)  \models \varphi'_1 \until_I \varphi'_2 \\ \hspace*{6mm}\text{ iff for some } \tau_j\in \tau_i+I, ( j ,\env) \models\varphi_2'\text{ and } ( k ,\env) \models \hspace*{6mm} \varphi_1', \forall i \le k<j \\
\hspace*{6mm}\text{ iff for some } \tau_j\in \tau_i+I, \rho(\varphi_2 , i,\env) \ge \hspace*{6mm} r \text{ and } \rho(\varphi_1 , k,\env) \ge r, \forall i \le k<j \\
\hspace*{6mm} \text{ iff } \displaystyle\max_{\tau_j \in \tau_i+I} \min \left( \rho(\varphi_2,   j, \env),\min_{\tau_k \in [\tau_i,\tau_j) }\rho(\varphi_1,   k, \env) \right) \ge r \\
\hspace*{6mm} \text{ iff }\rho(\varphi_1 \until_I \varphi_2 , i,\env) \ge r  
$\\
$
\bullet \text{ Case }(i,\env) \not \models \varphi'_1 \until_I \varphi'_2 \\
\hspace*{6mm} \text{ iff for all } \tau_j \in \tau_i + I, (j,\env) \not \models \varphi'_2 \text{ or there exists } k \in \hspace*{6mm}[i,j], ( k ,\env) \not \models \varphi_1' \\ 
\hspace*{6mm}\text{ iff for all } \tau_j \in \tau_i + I, \rho(\varphi_2 , i,\env) \le r \text{ or there exists } k \in \hspace*{6mm}[i,j], \rho(\varphi_1 , k,\env) \le r \\ 
\hspace*{6mm} \text{ iff } \displaystyle\max_{\tau_j \in \tau_i+I} \min \left( \rho(\varphi_2,   j, \env),\min_{\tau_k \in [\tau_i,\tau_j) }\rho(\varphi_1,   k, \env) \right) \le r \\
\hspace*{6mm} \text{ iff }\rho(\varphi_1 \until_I \varphi_2 , i,\env) \le r  \qed
$\\ 

We omit the case of the freeze operator since it just copies the values of its child in the syntax tree. \qed

\vspace{1cm}

\textbf{Lemma \ref{lemm} proof}: Given an \STLstar formula $\varphi$ and its associated syntax tree $\myast(\varphi)$, the robustness value $\rho(\varphi,0,\env)$ is the robustness value of the highest node in $\myast(\varphi)$. For each node in $\myast(\varphi)$, a robustness value of that node is obtained by applying $\max/\min$ operations to the robustness values of its children (look definition \ref{definition:Robustness}). This brings us to the starting point which is the leaves of $\myast(\varphi)$. If we look again at the definition of the quantitative semantics for the signal predicates and constraints, we can see that the robustness values at the leaves depend on $f_1$,$f_2$ and on the different signal values $\sigma_i^k, i\in [0,|\pi|-1], k \in [1,\dm]$. Thus, we are able to give an upper and lower bound on $\rho(\varphi,0,\env)$ by examining the different signal constraints and predicates in $\varphi$. \qed

\vspace{1cm}

\textbf{Proposition \ref{prop} proof}:
Given a signal constraint $\psi$ in $\varphi$ of the form $f_1(\overline{s}) \sim f_2(\overline{s_*})$, it takes $O(|\pi|^{|V|})$ to obtain the maximum and minimum possible value of $f_2(\env(s^{1}_*),...,\env(s^{\dm}_*))$ for all possible environments $\env$. Once we have these two values, we can obtain the maximum and minimum possible robustness values for $\psi$. We suppose we have a constant number of signal constraints in $\varphi$ and the interval $[a,b]$ will be composed of the maximum and minimum possible robustness values for any signal constraint or predicate in $\varphi$.
    Additionally, as we described our quantitative monitoring procedure, each new call of the monitoring algorithm divides the conservative range by 2. The initial interval $[a,b]$ can be divided to $\frac{b-a}{\epsilon}$ intervals each of them has width $\epsilon$. Thus, it will take $\lceil\log_2(\frac{b-a}{\epsilon})\rceil$ calls to reach a conservative range of size less or equal to $\epsilon$. \qed